\documentclass[aps,prb,twocolumn,superscriptaddress,longbibliography]{revtex4-1}

\usepackage[caption=false]{subfig}
\usepackage[colorlinks=true,citecolor=blue]{hyperref} % hyperreferences, use \autoref{} for referencing
\usepackage{graphicx}
\usepackage{physics}
\usepackage{bm}
\usepackage{amsmath}
\usepackage{comment}
\usepackage{amssymb}
\usepackage{qcircuit} % quantum circuits
\usepackage{blindtext}
\DeclareGraphicsExtensions{.png,.jpg,.eps}
\usepackage{xcolor}

\begin{document}
%%%%%%%%%%%%%%%%

%%%%%%%%%%%%%%%%%%%%%%%%%%
% Authors, Title, Abstract
%%%%%%%%%%%%%%%%%%%%%%%%%%

\author{Marcel Niedermeier}
\affiliation{Department of Applied Physics, Aalto University, 02150 Espoo, Finland}

\author{Marc Nairn}
\affiliation{Institut für Theoretische Physik, Universität Tübingen, 72076 Tübingen, Germany}

\author{Christian Flindt}
\affiliation{Department of Applied Physics, Aalto University, 02150 Espoo, Finland}
\affiliation{RIKEN Center for Quantum Computing, Wakoshi, Saitama 351-0198, Japan}

\author{Jose L. Lado}
\affiliation{Department of Applied Physics, Aalto University, 02150 Espoo, Finland}

\title{Quantum computing topological invariants of two-dimensional quantum matter}

\begin{abstract}
Quantum algorithms provide a potential strategy for solving computational problems that are intractable by classical means. Computing the topological invariants of topological matter is one central problem in research on quantum materials, and a variety of numerical approaches for this purpose have been developed. However, the complexity of quantum many-body Hamiltonians makes calculations of topological invariants challenging for interacting systems. Here, we present two quantum circuits for calculating Chern numbers of two-dimensional quantum matter on quantum computers. Both circuits combine a gate-based adiabatic time-evolution over the discretized Brillouin zone with particular phase estimation techniques. The first algorithm uses many qubits, and we analyze it using a tensor-network simulator of quantum circuits. The second circuit uses fewer qubits, and we implement it experimentally on a quantum computer based on superconducting qubits. Our results establish a method for computing topological invariants with quantum circuits, taking a step towards characterizing interacting topological quantum matter using quantum computers.
\end{abstract}

\date{\today}

\maketitle

\section{Introduction}

\begin{figure}[h!]
    \centering
    \includegraphics[width=\columnwidth]{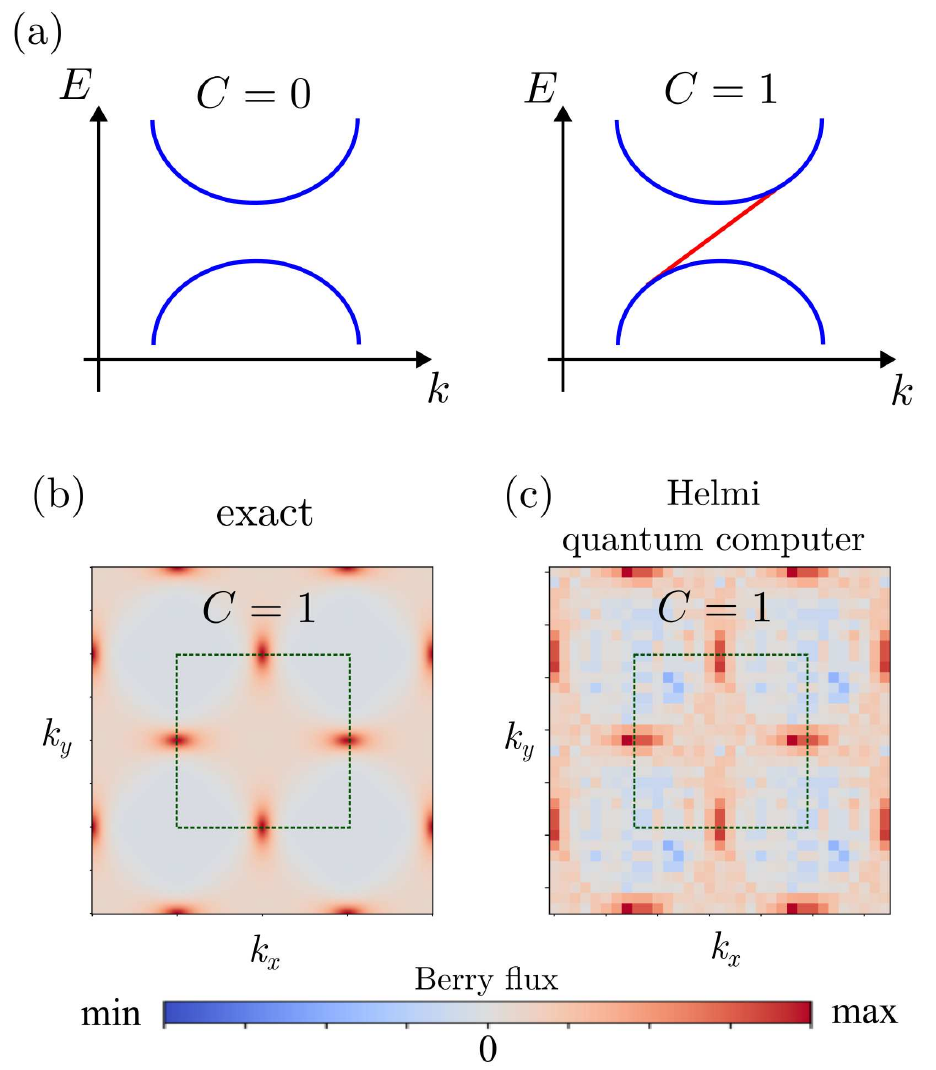}
    \caption{Quantum computations of topological invariants. (a)~Two-dimensional quantum matter can be characterized
    by the Chern number $C$. A non-zero Chern number implies that edge states exist as indicated by a red line. (b,c) A quantum circuit can be used to calculate the Berry flux and the Chern number. The two panels show a comparison between exact results for the Berry flux in the extended Brillouin zone and actual calculations on the quantum computer Helmi.}
    \label{fig:overview}
\end{figure}

Classifying topological phases of matter is a central problem in modern condensed matter physics.\cite{RevModPhys.83.1057,RevModPhys.82.3045,Moessner2021,asboth2016short,Berry_phase_effects_elec_prop, Ren2016}  In recent years, it has been realized that topological matter harbors a range of exotic phenomena, including the existence of chiral edge states\cite{RevModPhys.95.011002} and 
helical modes,\cite{Maciejko2011} as well as fractional excitations and Majorana fermions.\cite{Alicea2012}
Topological phases of matter may be characterized by topological invariants, such as the Berry phase in one dimension \cite{Berry_initial, PRLZak} and the Chern number in two dimensions, Fig.~\ref{fig:overview}(a).\cite{Haldane1988, quantum_spin_Hall_effect_Graphene} 
Ultimately, it is expected that certain topological excitations will make it possible to build a topological quantum computer that will be protected against the errors that limit current quantum processors.~\cite{Kitaev2001,Alicea2011,top_QC,Lahtinen2017} 

Computational methods have been successful in predicting the topological phases of matter that can be described by single-particle Hamiltonians.\cite{Fukui2005, Chern_number_without_integration} However, for strongly-correlated topological matter, where interactions may be important, the required computational resources scale exponentially with the system size, making it hard to predict their topological phase diagram. Variational eigensolvers implemented on quantum computers provide one promising strategy to characterize strongly-correlated quantum matter.\cite{QC_many_body_physics} As such, 
quantum circuits for computing topological
invariants of quantum matter have become a vibrant area of research.\cite{Berry_QC, Semeghini2021, Xiao2021determiningquantum, Xiao2023robustmeasurementof, QC_alg_many_body_top_invariants, dig_simulation_top_matter_programmable_quantum_processors, crossing_top_phase_transition_QC, identification_sym_protected_top_states_QC, measurement_ent_spectrum_sym_prot_top_state, Roushan2014, observing_top_invariants_quantum_walks, winding_number_large_scale_chiral_quantum_walk, top_inv_non_unitary_discrete_time_quantum_walks, var_QC_corr_top_phases,PhysRevResearch.3.023244}

\begin{figure*}[t!]
    \centering
    \includegraphics[width=0.99\textwidth]{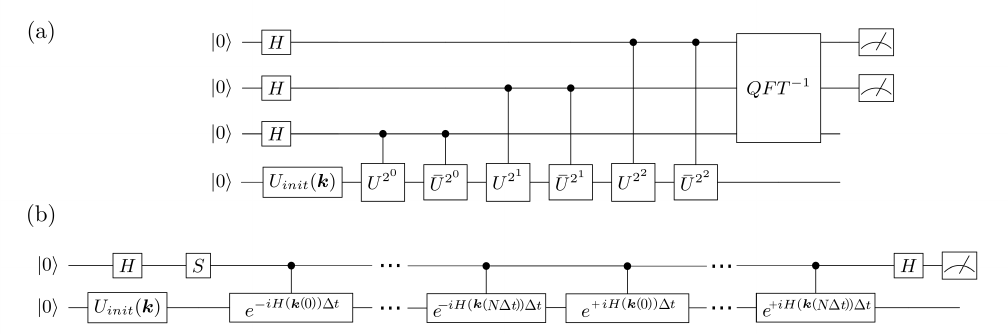}
    \caption{Quantum circuits for calculating Berry phases and fluxes. (a) The operator $U_{\text{init}}(\boldsymbol{k})$ initializes the lowest qubit followed by consecutive applications of the time-evolution operators, $U=U(T,0)$ and $\bar{U}=\bar U(T,0)$, defined in connection with Eq.~(\ref{eq:berryunit}). The first three qubits are used to read out the Berry phase with the quantum phase estimation algorithm. Hadamard gates are denoted by $H$. The inverse quantum Fourier transform is denoted as QFT$^{-1}$. (b)~Quantum circuit with two qubits for reading out the Berry fluxes using a Hadamard test. Here, we show the adiabatic loop explicitly. Adding or omitting the $S$-gate determines if the Berry flux should be determined using a sine or cosine function.}
    \label{fig:QC}
\end{figure*}

In this work, we implement two quantum circuits for calculating the Chern number of two-dimensional quantum matter. The first circuit uses the quantum phase estimation algorithm to extract the Chern number from the
winding of the Wannier center across the unit cell. This circuit typically requires many qubits, and we analyze it using a tensor-network simulator of quantum circuits. The second circuit exploits that the Chern number can be 
calculated by summing up the local Berry fluxes in momentum space, which can be determined using a Hadamard test algorithm. Building on the quantum algorithms from Refs.~\onlinecite{Berry_QC, Xiao2023robustmeasurementof}, we employ an adiabatic quantum state evolution along a two-dimensional path, which has the advantage of requiring only a single ground state preparation step per plaquette. The circuit only requires a single auxiliary qubit, and we implement it using the quantum computer Helmi.\cite{Helmi_technical} As seen in Fig.~\ref{fig:overview}(b,c), we find good agreement between our quantum computations and exact results for the Berry flux. As such, we show how quantum circuits make it possible to determine topological invariants of quantum matter, providing a promising starting point for characterizing strongly-correlated quantum matter using quantum computers.

The rest of this paper is organized as follows. In Sec.~\ref{Sec_methods}, we introduce the two quantum circuits that we use to compute the Berry phase and the Chern numbers. We also describe the quantum computer and the tensor-network simulator of quantum circuits that we use for our calculations. In Sec.~\ref{Sec_results}, we present our results for the topological classification of the Qi-Wu-Zhang (QWZ) model and the Haldane model. 
Finally, in Sec.~\ref{Sec_conclusion}, we present our conclusions together with a brief outlook on possible avenues for future developments. Several technical details are described in two appendices.

\section{Methods}
\label{Sec_methods}

\subsection{Berry phase calculations}
We first show how the Berry phase emerges from a Bloch Hamiltonian  $H(\boldsymbol{k})$ that depends on the momentum~$\boldsymbol{k}$. For each momentum, the Hamiltonian has a complete set of eigenstates denoted by $\{ \ket{n(\boldsymbol{k})} \}$ with
\begin{equation}
    H(\boldsymbol{k}) \ket{n(\boldsymbol{k})} = E_n(\boldsymbol{k}) \ket{n(\boldsymbol{k})}. 
\end{equation}
We now initialize the system in the eigenstate $\ket{n(\boldsymbol{k})}$ and let it evolve while changing the momentum in a closed loop from its initial value $\boldsymbol{k}(t=0)=\boldsymbol{k}$ to the final value $\boldsymbol{k}(t=T)=\boldsymbol{k}$. Moreover, we change the Hamiltonian slowly enough that the system remains in its instantaneous eigenstate $\ket{n(\boldsymbol{k}(t))}$ at all times, up to an overall phase factor. We can therefore write
\begin{equation}
    U(T, 0) \ket{n(\boldsymbol{k})} =  e^{i (\Theta_B-\Theta_D)} \ket{n(\boldsymbol{k})}, 
\end{equation}
where 
\begin{equation}
U(T, 0)=\mathcal{T}\{e^{-i\int_0^TdtH(\boldsymbol{k}(t))}\}
    \label{eq:berryU}
\end{equation}
is the time-evolution operator. According to the adiabatic theorem, the total phase that is picked up is given by the difference between the Berry phase,
\begin{equation}
\Theta_B = i \int_0^T dt  \langle n(\boldsymbol{k}(t))|\partial_{t}|  n(\boldsymbol{k}(t))\rangle,
\label{eq:ber_phas}
\end{equation}
and the dynamical phase,
\begin{equation}
    \Theta_D = \int_0^T dt E_n(\boldsymbol{k}(t)).
    \label{eq:dyn_phas}
\end{equation}
If we condition $U(T, 0)$  on an auxiliary qubit, the overall phase $\Theta_B-\Theta_D$ can be measured using a Hadamard test~\cite{Hadamard_test} or the quantum phase estimation algorithm.\cite{Kitaev_QPE, nielsen_chuang_2010} 

In the following, we wish to determine the Berry phase~$\Theta_B$ only. To this end, we let the system evolve through the same closed loop in momentum space one more time. However, we now let the system evolve backwards in time and denote the corresponding time-evolution operator by $\bar{U}(T,0)$. It is defined as in Eq.~(\ref{eq:berryU}), however, with the opposite sign in the exponent. One may equivalently think of this unitary operator as propagating the system forward in time, but with the opposite sign of the Hamiltonian. We then easily see that the dynamical phase in Eq.~(\ref{eq:dyn_phas}) changes sign, since the eigenvalues change sign, while the Berry phase remains the same, since the eigenvectors are unchanged. After having traversed the closed loop twice, we therefore find
\begin{equation}
    \bar{U}(T,0) U(T, 0) \ket{n(\boldsymbol{k})} =  e^{i 2\Theta_B} \ket{n(\boldsymbol{k})}, 
    \label{eq:berryunit}
\end{equation}
which allows us to determine the Berry phase only. 

Figure~\ref{fig:QC} shows the two quantum circuits for determining the Berry phase.  The first circuit extracts the Berry phase using a quantum phase estimation algorithm. Typically, this circuit requires many qubits, so in the following we only simulate it using the tensor-network simulator that we describe below. The second circuit instead uses a Hadamard test to extract the Berry phase. Details of the Hadamard test are provided in App.~\ref{app:hadamard}. This circuit requires fewer qubits, and we can implement it experimentally on a current quantum computer. In both circuits, we realize the time-evolution operators by a sequence of small time steps,~\cite{Berry_QC} such that
\begin{equation}
U(T, 0)\simeq 
    \prod_{j=1}^{N} e^{-iH(\boldsymbol{k}(j \Delta t))\Delta t},
\end{equation}
where $N$ is the number of time steps of length $\Delta t$ with $N \Delta t = T$, with an equivalent expression for $\bar{U}(T,0)$.

To check the output of the quantum circuits, we can also calculate the exact values of the Berry phase using the conventional Wilson-loop algorithm.\cite{PhysRevB.56.12847,PhysRevLett.102.107603,Wu2018,PhysRevB.84.075119,PhysRevB.95.075146,PhysRevB.83.235401}
Thus, we discretize a closed loop in momentum space as $\{ \boldsymbol{k}_1, \boldsymbol{k}_2, ..., \boldsymbol{k}_N, \boldsymbol{k}_1 \}$, assuming a non-degenerate ground state, and the Berry phase  then reads 
\begin{equation}
\Theta_B = \text{arg} 
\left [ M_{\boldsymbol{k}_1,\boldsymbol{k}_2}
M_{\boldsymbol{k}_2,\boldsymbol{k}_3}
... M_{\boldsymbol{k}_N,\boldsymbol{k}_1} 
\right ],
\label{eq:Wilson-loop}
\end{equation}
where we have defined the overlaps of eigenstates as
\begin{equation}
M_{\boldsymbol{k}_\alpha,\boldsymbol{k}_\beta} = \braket{n(\boldsymbol{k}_\alpha)}{n(\boldsymbol{k}_\beta)}.
\end{equation}

\subsection{Chern number calculations}

In the following, we use the quantum circuits
from above to compute the Chern number
of two-dimensional systems. For the circuit in Fig.~\ref{fig:QC}(a), we use the fact that the Berry phase directly yields the positions of the hybrid Wannier centers, defined below, and we can then determine the Chern number directly from their winding across the unit cell.\cite{PhysRevB.56.12847,PhysRevLett.102.107603,Wu2018,PhysRevB.84.075119,PhysRevB.95.075146,PhysRevB.83.235401} For the circuit in Fig.~\ref{fig:QC}(b), we extract the Chern number from a direct calculation of the Berry flux in momentum
space.\cite{RevModPhys.95.011002}
For the circuit in Fig.~\ref{fig:QC}(a), we need the quantum phase estimation algorithm to accurately determine the Berry phase, which is sizeable.  By contrast, the circuit in Fig.~\ref{fig:QC}(b) only relies on the extraction of small Berry phases in momentum space, and that can be done using a Hadamard test.

For the circuit in Fig.~\ref{fig:QC}(a), we first show how the Chern number can be extracted from the winding number of a hybrid Wannier function. For a set Bloch functions~$\ket{n(\boldsymbol{k})}$, the Wannier center is defined as
\begin{equation}
    \boldsymbol{R}_{W} = \braket{w_n(\boldsymbol{0})}{\boldsymbol{R} \, \vert w_n(\boldsymbol{0})}
\end{equation} 
in terms of the Wannier functions
\begin{equation}
    \ket{w_n(\boldsymbol{R})} = \frac{1}{2 \pi} \int_{BZ} d^2\boldsymbol{k} \, e^{i \boldsymbol{k} \cdot \boldsymbol{R} } \ket{n(\boldsymbol{k})}
\end{equation} 
evaluated at $\boldsymbol{R}=\boldsymbol{0}$. We can also define hybrid Wannier functions by keeping one of the momenta fixed and integrating over the other momentum. The center of the hybrid Wannier function in the~$x$-coordinate as a function of~$k_y$ is then given by
\begin{equation}
X_{W}(k_y) = 
    i\int_{-\pi}^{\pi} d k_x 
     \braket{n(\boldsymbol{k})}{ \partial_{k_x} \vert n(\boldsymbol{k})}/2\pi.
\end{equation}
By comparing this expression with Eq.~(\ref{eq:ber_phas}), we see that the Wannier center is equal to the Berry phase, $X_{W}(k_y)=
{\Theta_B} (k_y)$, through a change of variables. 
Thus, the hybrid Wannier center can be obtained from the momentum-dependent Berry phase $\Theta_B(k_y)$ by performing the adiabatic evolution over $k_x$ from~$-\pi$ to~$\pi$. By counting the number of times the center $X_W$ winds around the unit cell, 
we find the Chern number.\cite{PhysRevB.56.12847,PhysRevLett.102.107603,Wu2018,PhysRevB.84.075119,PhysRevB.95.075146,PhysRevB.83.235401}

\begin{figure*}
    \centering
    \includegraphics[width=0.99\textwidth]{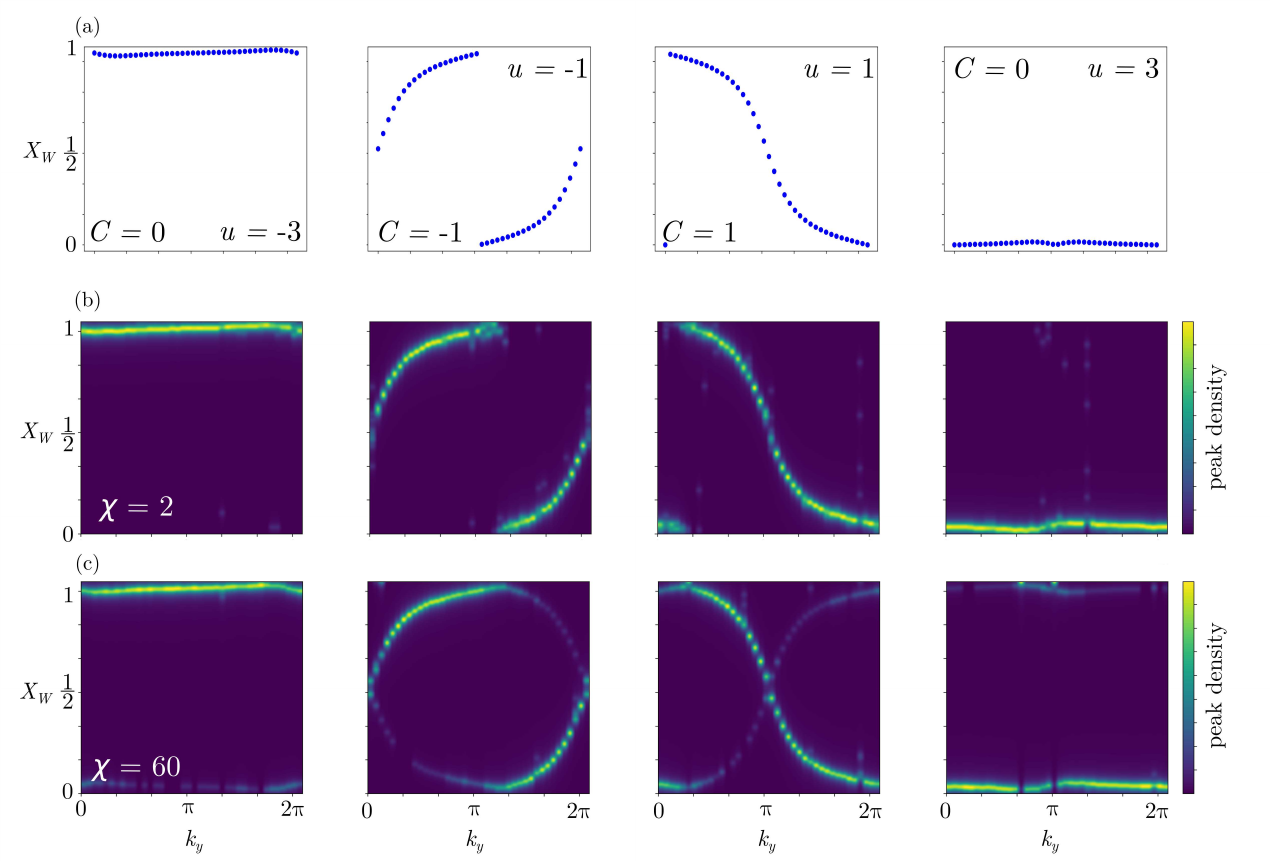}
    \caption{Wannier centers of the QWZ model. Results are shown for different values of the on-site potential, $u = -3, -1, 1, 3$, corresponding to different topological phases. (a) Exact calculations of the Wannier centers $X_W$ together with the Chern numbers that are extracted from the winding across the Brillouin zone. (b) Tensor-network simulation of the quantum circuit in Fig.~\ref{fig:QC}(a) with a maximum bond dimension of $\chi = 2$. (c) Similar simulations with a maximum bond dimension of $\chi = 60$.}
    \label{fig:QWZ_Berry_DOS_Chern_number}
\end{figure*}

For the circuit in Fig.~\ref{fig:QC}(b), we calculate the Chern number
directly by integrating over the Berry flux.
The Chern number is defined as the integral 
\begin{equation}
    C = \frac{1}{2\pi} \int_{BZ} d^2\boldsymbol{k} \Omega(\boldsymbol{k}),
\end{equation}
where the Berry curvature,
\begin{equation}
\Omega (\boldsymbol{k})
=
\partial_{k_x} A_{k_y}
-
\partial_{k_y} A_{k_x},
\end{equation}
is defined as the curl of the Berry connection 
\begin{equation}
\boldsymbol{A} = i \langle n(\boldsymbol{k})| \nabla_{\boldsymbol{k}}| n(\boldsymbol{k})\rangle.
\end{equation}

To calculate the Chern number, we decompose the Brillouin zone into $N \times N$ plaquettes of size $(\Delta k)^2$, such that $\Delta k = 2\pi/N$. The Chern number is then 
\begin{equation}
    C \simeq\frac{1}{2\pi} \sum_{i, j = 1}^{N} \Omega_{ij},
    \label{eq:chern_appr}
\end{equation}
where $\Omega_{ij}$ is the Berry flux through the plaquette at position $(i, j)$ in the Brillouin zone. This flux can be computed as the Berry phase accumulated during an adiabatic evolution around the plaquette~\cite{Fukui2005} given by the loop
\begin{equation}
\begin{split}
    (k_x, k_y) &\rightarrow (k_x + \Delta k, k_y) \rightarrow (k_x + \Delta k, k_y + \Delta k) \\
    &\rightarrow (k_x, k_y + \Delta k) \rightarrow (k_x, k_y).
\end{split}
\end{equation}
Repeating this procedure for every plaquette and summing up the resulting Berry fluxes, we then find the Chern number according to Eq.~(\ref{eq:chern_appr}). For small plaquettes, the associated Berry flux is small enough, typically below $10^{-2}$, that we can apply the Hadamard test algorithm to read out the Berry phase at the end of each adiabatic evolution. For this approach, the preparation of the eigenstate $\ket{n(\boldsymbol{k})}$ is only required once per plaquette. For our calculations, we prepare the initial state $\ket{n(\boldsymbol{k})}$ using the similarity transformation $U_{\text{init}}(\boldsymbol{k})$ that diagonalizes the Hamiltonian. By contrast, more complicated systems might require a ground-state preparation based on a variational quantum eigensolver for instance.\cite{Tilly2022}

\begin{figure*}
    \centering
    \includegraphics[width=0.99\textwidth]{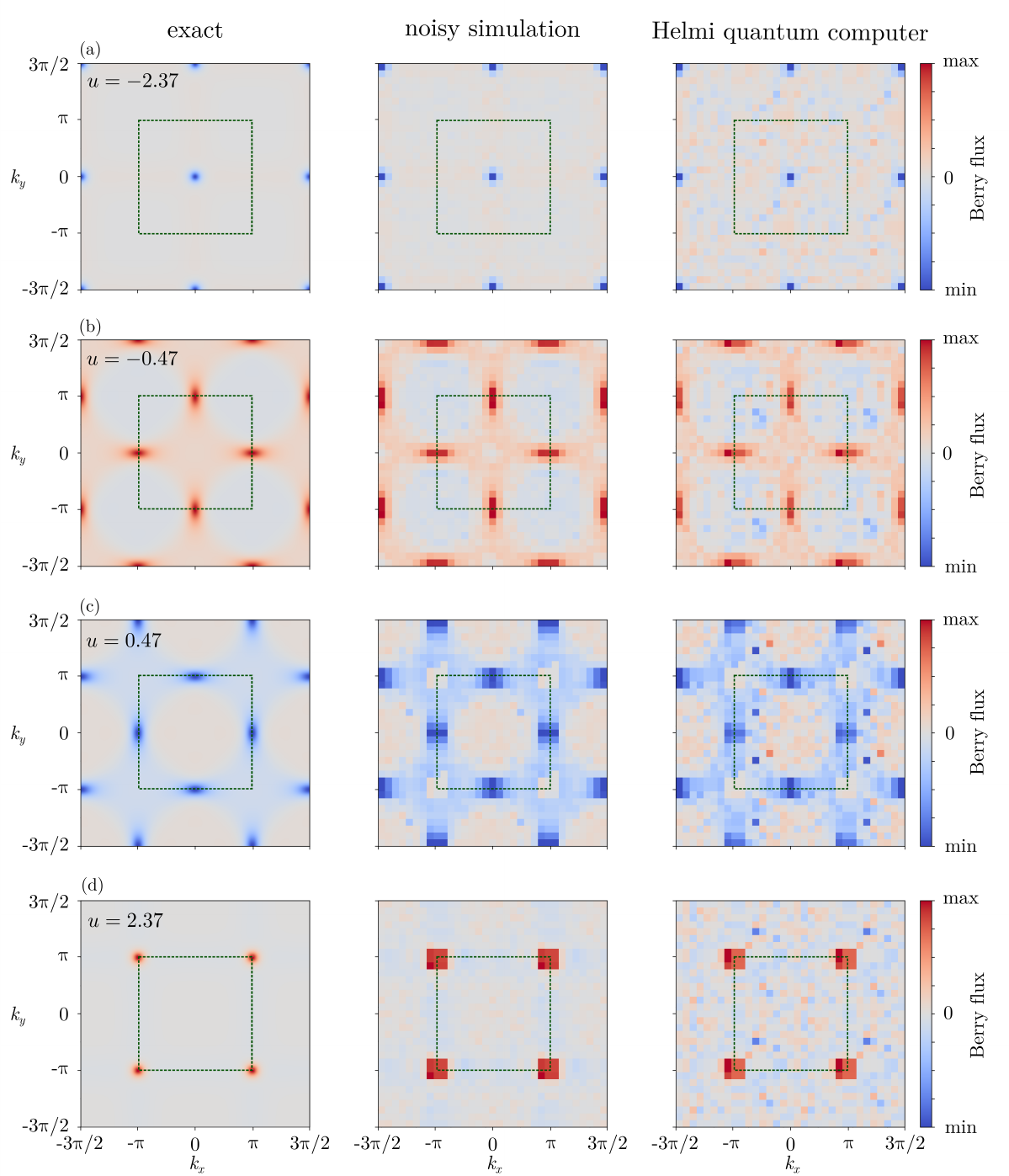}
    \caption{Berry flux of the QWZ model. The extended Brillouin zone is divided into $15 \times 15$ plaquettes, while the first Brillouin zone is indicated by a dashed square. From left to right, we show exact results, results from noisy simulations, and results from Helmi. The rows correspond to different on-site potentials $u$. These results are used to find the Chern numbers in Fig.~\ref{fig:Chern_QWZ_Helmi}.}
    \label{fig:berry_flux}
\end{figure*}

\subsection{Quantum computations and tensor-network simulations}

We implement our quantum circuit calculations on the Helmi quantum computer.\cite{Helmi_technical} Helmi is a 5-qubit quantum computer based on superconducting qubits, which are connected in a star-shaped topology. Its native quantum gates are the controlled-$Z$ gate and the one-qubit phased rotation gate around the $x$-axis, and it can be steered with standard Qiskit instructions. Helmi's $T_1$ and $T_2$ times are $35.7$~$\mu$s and $17.4$~$\mu$s with a one- and two-qubit native gate application time of $120$~ns, allowing for a circuit depth of about $170$ native gates before qubit decoherence becomes an issue. Typical one- and two-qubit gate fidelities are $99.6$~$\%$ and $96.1$~$\%$, respectively. \footnote{All of these values are subject to daily fluctuations and are averaged over all five qubits~\cite{Helmi_VTT}} An important parameter for our circuit implementations is the optimization level of the transpilation to the quantum processor. Here we always use the highest possible level, which is level $3$. We also perform noisy simulations of Helmi with the fake backend \texttt{FakeAdonis()}.

The circuit in Fig.~\ref{fig:QC}(a) produces large Berry phases, and we therefore need to use the quantum phase estimation algorithm, instead of a Hadamard test. For this reason, the circuit requires more qubits, and it cannot be implemented on the Helmi quantum computer currently. Instead, we simulate this quantum circuit using matrix product states, which are a subclass of tensor networks.\cite{Schollwock_DMRG} 

Using matrix product states, a generic quantum state
\begin{equation}
\ket{\Psi} = 
\sum_{\boldsymbol{\sigma}} \Gamma^{\boldsymbol{s}}|\boldsymbol{s}\rangle
\end{equation}
can be represented by the tensor train
\begin{equation}
     \Gamma^{\boldsymbol{s}} = \sum_{{\boldsymbol{s}}} [A_1]_{\chi_1}^{s_1} [A_2]_{\chi_1 \chi_2}^{s_2} ... [A_N]_{\chi_{N-1}}^{s_N},
\end{equation}
where $\{\ket{\boldsymbol{s}}\}$ are the computational basis states. The size of each tensor, $A_i$, is controlled by the bond dimension $\chi_i$. Thus, by adjusting the maximum bond dimension, an upper bound on the entanglement in a quantum state can be imposed. As such, we can simulate the generic loss of entanglement in a quantum computer by reducing the bond dimension.\cite{TN_quantum_sim_lim_ent, PRX_what_limits, Grover_no_quantum_advantage, DMRG_QC, Shor_alg_MPS, Shor_alg_MPS_optim, Woolfe2017} Here, our tensor-network simulations are performed using the \texttt{ITensors} package,\cite{Itensors} interfaced to our publicly available quantum circuit simulation package called Quantunity.\cite{QSim_github}

\section{Results}
\label{Sec_results}

\subsection{Qi-Wu-Zhang model}

\begin{figure}
    \centering
    \includegraphics[width=0.99\columnwidth]{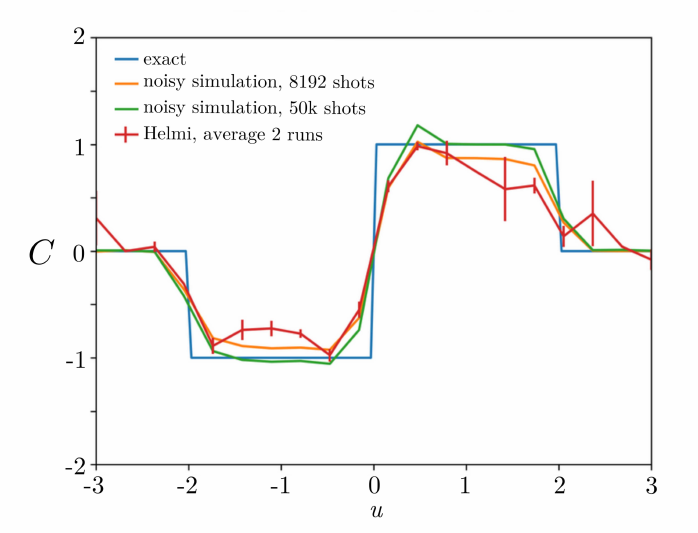}
    \caption{Chern number of the QWZ model. The Chern number is shown as a function of the on-site potential $u$. Results from Helmi are compared with exact results and simulations. }
    \label{fig:Chern_QWZ_Helmi}
\end{figure}

The Qi-Wu-Zhang (QWZ) model describes a two-dimensional topological insulator with a non-vanishing Chern number.\cite{QWZ2006} The bulk Hamiltonian reads
\begin{equation}
\begin{split}
    H_\text{QWZ}(\boldsymbol{k}) &= \sin(k_x)\sigma_x + \sin(k_y)\sigma_y \\
    &+\left[u+\cos(k_x)+\cos(k_y)\right]\sigma_z,
\end{split}
\end{equation}
where the on-site potential $u$ controls the topological phase of the system. Indeed, the energy spectrum splits into two bands with gap-closings at $u=2, 0, -2$. Between the gap-closings, one finds non-trivial topological phases with ${C}=-1$ for $-2<u<0$ and ${C}=1$ for $0<u<2$, and ${C}=0$ otherwise.\cite{asboth2016short}

In Fig.~\ref{fig:QWZ_Berry_DOS_Chern_number}, we show the hybrid Wannier center of the QWZ model for different on-site potentials together with the Chern numbers we extract. Here, we can simplify the numerical calculations by using the fact that the bulk Hamiltonian of the QWZ model is mirror symmetric, i.e. that $E_0(k_x, k_y) = E_0(- k_x, k_y)$ for $k_x, k_y \in [0, 2\pi]$. It is therefore sufficient to change the sign of the time evolution in the second half of the adiabatic evolution, for $k_x > \pi$, as the symmetry of the energy bands guarantees an exact cancellation of the dynamical phase for each pair of momenta $k_x$ and $-k_x$. Figure~\ref{fig:QWZ_Berry_DOS_Chern_number}(a) shows the exact results that we obtain using the Wilson-loop algorithm in Eq.~(\ref{eq:Wilson-loop}). In Figs.~\ref{fig:QWZ_Berry_DOS_Chern_number}(b,c), we show results obtained with our tensor-network simulations of the quantum circuit in Fig.~\ref{fig:QC}(a) with a 12-qubit quantum phase estimation algorithm (1 qubit for the ground state and 11 qubits in the work register, out of which 9 are measured). Each half of the adiabatic evolution proceeds in $N_k=100$ increments, i.e. the momentum $k_x$ increases in steps of $\pi/100$. The time increment parameter $dt$ is chose to reach a maximum time of $T=10$, i.e. $dt = T/N_k$. We show the probability density of the hybrid Wannier center $P(X_W)$ obtained from the quantum phase estimation algorithm,\footnote{
The probability density is computed as
$
    P(X_W) = \sum_j \frac{\varepsilon}{\varepsilon^2 + (X_W - X_W^j)^2},
$
where $X^j_W=\theta_B^j$ are the Berry phases obtained from the quantum phase estimation algorithm, and the small parameter $\varepsilon= 0.1$ broadens the peaks.}
using different bond dimensions in the two panels. In both cases, we find good agreement with the exact results in Figs.~\ref{fig:QWZ_Berry_DOS_Chern_number}(a), and the Chern number can accurately be extracted from the winding of the Berry phase across the Brillouin zone. We find that the results do not depend strongly on the bond dimension, indicating that the algorithm may not be very susceptible to the loss of entanglement. For bond dimensions less than $2$, we find that the extraction of the Wannier center density in the quantum phase estimation is not accurate.

In Fig.~\ref{fig:berry_flux}, we show calculations of the Berry flux using the quantum circuit in Fig.~\ref{fig:QC}(b). For this circuit, quantum calculations are performed using Helmi. For the sake of comparison, we also show exact results from the Wilson-loop algorithm as well as simulations of Helmi using the \texttt{FakeAdonis()} backend. To calculate the Berry flux, we discretize the Brillouin zone into a grid of $15 \times 15$ plaquettes. Employing two evolution steps per link, meaning an adiabatic transformation by $2\pi/30$ degrees per time-evolution step, we obtain a circuit depth of 6 elementary single-site and 2 two-site gates per plaquette at the transpilation level 3. The results are then read out with $8192$ shots. Figure~\ref{fig:berry_flux} shows the Berry flux across the extended Brillouin zone, with the first Brillouin zone, $[-\pi, \pi]\cross [-\pi, \pi]$, highlighted by the dashed squares. Exact results are shown in the left column, while noisy simulations of the
quantum circuit are presented in the middle column. In the right column, we show the results obtained experimentally with the Helmi quantum computer. Each row corresponds to different values of the on-site potential, which determines the topological phase of the system. Generally, we observe a good agreement between the calculations based on Helmi and both the exact results and those based on noisy simulations of Helmi. Importantly, from these calculations, we can extract the corresponding Chern numbers.

Figure~\ref{fig:Chern_QWZ_Helmi} shows the Chern number obtained by integrating the Berry fluxes in Fig.~\ref{fig:berry_flux} over the unit cell. The results from Helmi are averaged over two runs, and the error bars indicate the difference between the largest and the smallest values. We also show the exact results as well as results obtained from noisy simulations of Helmi. The results from Helmi clearly allow us to distinguish the different topological phases of the system. However, the results from Helmi and the noisy simulations are often smaller than the exact Chern number, which is an integer. However, we see that this issue can be resolved by increasing the number of measurements for the noisy simulations. This observation is consistent with the fact that the Hadamard test is very sensitive to the measurement precision, since the measurement frequencies enter into an inverse trigonometric function.

Finally, we note that the results discussed above accurately reflect the performance of a currently available superconducting quantum processor. As such, error mitigation techniques are not necessarily required to perform a topological classification in this case. To summarize, we find that both circuits in Fig.~\ref{fig:QC} make it possible to determine the Chern numbers of the QWZ model with error margins that generally are small.

\subsection{Haldane model}

\begin{figure}
    \centering
    \includegraphics[width=0.99\columnwidth]{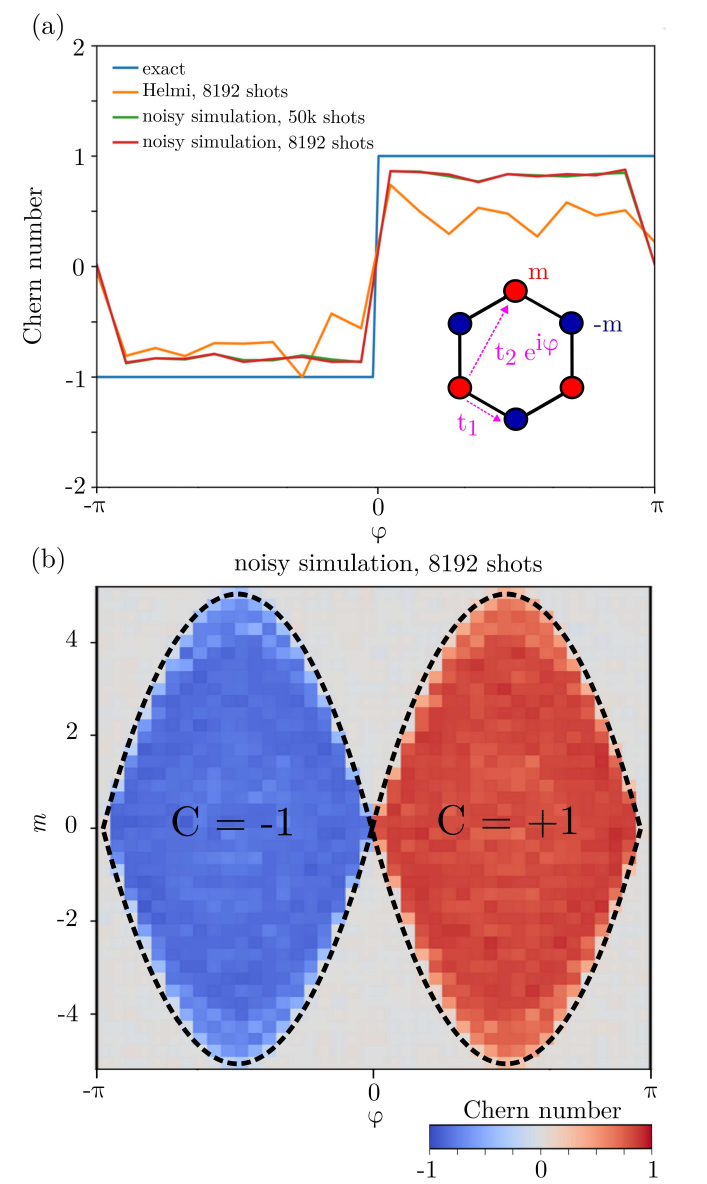}
    \caption{Topological classification of the Haldane model.~(a)~Chern number as a function of the phase with zero mass. Results obtained from Helmi are compared with the exact results and noisy simulations. (b) Full topological phase diagram obtained by noisy simulations of Helmi. The dashed lines indicate the exact phase boundaries given by Eq.~(\ref{eq:haldane_bound}).}
    \label{fig:Chern_Haldane_Helmi_heatmap}
\end{figure}

As our second application, we consider the topological classification of the Haldane model. The Haldane model describes a two-dimensional Chern insulator on a honeycomb lattice. As illustrated in the inset of Fig.~\ref{fig:Chern_Haldane_Helmi_heatmap}(a), the model involves nearest-neighbor and next-nearest-neighbor hopping with amplitudes $t_1$ and $t_2 e^{i\varphi}$, where~$\varphi$ is a tunable phase. The bulk Hamiltonian reads
\begin{equation}
    H(\boldsymbol{k}) = \boldsymbol{d}(\boldsymbol{k}) \cdot \boldsymbol{\sigma}, 
\end{equation}
where the vector $\boldsymbol{\sigma}$ contains the Pauli matrices. The three components of the vector $\boldsymbol{d}(\boldsymbol{k})$ are
\begin{equation}
\begin{split}
d^x(\boldsymbol{k}) = & t_1 [\cos(\boldsymbol{k}\cdot \boldsymbol{a}_1) + \cos(\boldsymbol{k}\cdot \boldsymbol{a}_2) + 1], \\
d^y(\boldsymbol{k}) = & t_1 [\sin(\boldsymbol{k}\cdot \boldsymbol{a}_1) + \sin(\boldsymbol{k}\cdot \boldsymbol{a}_2)], \\
d^z(\boldsymbol{k}) = & m + 2t_2 \sin(\varphi) [\sin(\boldsymbol{k}\cdot \boldsymbol{a}_1)-\sin(\boldsymbol{k}\cdot \boldsymbol{a}_2)  \\ & - \sin(\boldsymbol{k}\cdot (\boldsymbol{a}_1 - \boldsymbol{a}_2))],
\end{split}
\end{equation}
where the mass term $m$ in $d^z(\boldsymbol{k})$ breaks time-reversal symmetry. The unit cell is spanned by the vectors
$\boldsymbol{a}_1 = (\sqrt{3}/2, 1/2) \quad\mathrm{and}\quad\boldsymbol{a}_2 = (\sqrt{3}/2, -1/2)$,
and the topological phase is controlled by the values of $m$ and $\varphi$. In the following, we focus on the case $t_1 = t_2 = 1$, where the system exhibits two topologically nontrivial phases. Specifically, the expected Chern numbers are
\begin{equation}
\begin{split}
{C}=-1, &\quad |m|<3\sqrt{3}|\sin(\varphi)|,\,\,  -\pi<\varphi<0, \\
{C}=+1, &\quad  |m|<3\sqrt{3}|\sin(\varphi)|,\,\, \phantom{-}0<\varphi<\pi,\\
{C}=\phantom{+}0, &\quad\text{otherwise (for } m\ne 0).
\end{split}
\label{eq:haldane_bound}
\end{equation}

In Fig.~\ref{fig:Chern_Haldane_Helmi_heatmap},
we show  the topological phase diagram
of the Haldane model computed with the quantum circuit in Fig.~\ref{fig:QC}(b) using a $15 \times 15$ grid in the Brillouin zone. In Fig.~\ref{fig:Chern_Haldane_Helmi_heatmap}(a) we show the Chern numbers obtained with Helmi as a function of the tunable phase and a zero mass. For the sake of comparison, we also show the exact results as well as noisy simulations. Also in this case, we observe a good qualitative agreement between the quantum computations and the exact results. The fact that the quantum simulations yield Chern numbers that are not exactly integers can be attributed to a slightly non-adiabatic time evolution. A higher accuracy would require more time steps, but also deeper quantum circuits, which would increase the net effective quantum error.
{Any deviation from the instantaneous ground state in the adiabatic evolution gives rise to a contribution from the excited state, representing the conduction band. As the conduction band has opposite Berry curvature, the resulting Chern number is systematically smaller in magnitude than in the perfect case. This phenomenology is analogous to the effect of finite temperature in an exact calculation of the Chern number.} Nevertheless, by rounding off to the closest integer, the results clearly allow us to distinguish between the topologically trivial and non-trivial phases. Finally, in Fig.~\ref{fig:Chern_Haldane_Helmi_heatmap}(b), we show the complete topological phase diagram of the Haldane model obtained with a noisy simulation. Here, we also observe a good agreement with the exact phase boundaries given by Eq.~(\ref{eq:haldane_bound}), shown with dashed lines.

\section{Conclusions}
\label{Sec_conclusion}
We have presented quantum computations of Chern numbers for two-dimensional quantum matter. To this end, we have implemented two quantum circuits that are based on the adiabatic evolution of a quantum state in momentum space. The first circuit uses the quantum phase estimation algorithm to determine the Berry phase. We can thereby extract the winding of the corresponding Wannier center across the unit cell, which directly yields the Chern number. This circuit requires many qubits to reach a high precision, and we have therefore simulated it using our tensor-network simulator of quantum circuits. The second circuit, by contrast, relies on calculations of the Berry flux through a grid of plaquettes in the Brillouin zone. The Berry flux associated with each plaquette is small enough that it can be read out using a Hadamard test. For this reason, the number of qubits in this circuit is small enough that it can be implemented on the quantum computer Helmi.  As specific applications, we have considered the topological classification of the  Qi-Wu-Zhang (QWZ) model and the Haldane model. For the QWZ model, we made use of both quantum circuits and found that the first circuit, based on the quantum phase estimation algorithm, can be implemented  
with a modest level of entanglement. With the second circuit, we found that the topological phases can be correctly identified using Helmi. The same conclusion was reached when calculating the topological phase diagram of the Haldane model using Helmi. For the second circuit, we found that it works well with a small number of evolution steps per link. The quantum circuit is therefore rather shallow and requires fewer qubits than the first circuit.
Our approach can be generalized to other topological invariants and may be applied to other systems such as $Z_2$ topological insulators or topological semimetals. Ultimately, the circuits may be implemented for interacting many-body Hamiltonians, where the many-body ground state is obtained via a variational eigensolver. A potential extension to interacting problems is outlined in App.~\ref{app:inter}. As such, our work provides a path towards classifying interacting quantum matter on quantum computers.

\section{Acknowledgments}

We acknowledge the financial support from the Finnish Quantum Flagship,
InstituteQ, the Jane and Aatos Erkko Foundation, the Aalto Science Institute, 
the Research Council of Finland through the Finnish Centre of Excellence in Quantum Technology (352925) and projects No.~331342 and 358088, and the Japan Society for the Promotion of Science through an Invitational Fellowship for Research in Japan. We acknowledge the computational resources provided by
the Aalto Science-IT project, CSC and the FiQCI quantum computing infrastructure.

\appendix

\section{Hadamard test scheme}
\label{app:hadamard}

\begin{figure}[t]
    \centering
    \includegraphics[width=0.95\columnwidth]{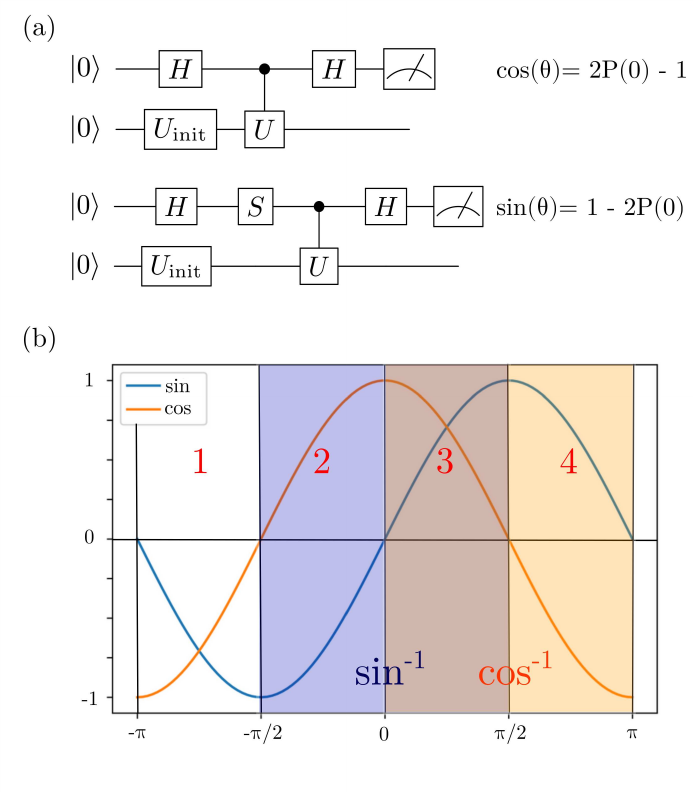}
    \caption{Hadamard test scheme. (a) Quantum circuits for finding the phase $\theta$ of the eigenvalue of the quantum state~$\ket{\psi}$, which is an eigenstate of $U$, which is prepared by $ U_{\text{init}}$. The eigenvalue is kicked back onto the upper, auxiliary qubit line. (b) In order to reconstruct phases on the interval $[-\pi, \pi]$, a combination of $\sin^{-1}$ and $\cos^{-1}$ has to be used.}
    \label{fig:Hadamard_test}
\end{figure}

The Hadamard test scheme measures the phase of a quantum state in a quantum circuit. Let $\ket{\psi}$ be an eigenstate of the operator $U$, which we assume is prepared by the operator $U_{\text{init}}$. We are interested in finding the phase~$\theta$, which determines the eigenvalue of $\ket{\psi}$ as
\begin{equation}
    U \ket{\psi} = e^{i\theta} \ket{\psi}.
\end{equation}
Here, we assume that $\theta \in [-\pi, \pi]$. The two quantum circuits in Fig.~\ref{fig:Hadamard_test} make use of a phase kickback by conditioning the operator $U$ on an auxiliary, upper qubit line. Measuring this qubit, one obtains the phase $\theta$ from the  probabilities found by repeated measurements as
\begin{equation}
    \cos(\theta) = 2P_{\text{cos}}(0) - 1, \quad 
    \sin(\theta) = 1 - 2P_{\text{sin}}(0). 
\end{equation}
To resolve phases $\theta$ in the entire interval $[-\pi, \pi]$, both circuits have to be executed, as the corresponding inverse functions $\sin^{-1}$ and $\cos^{-1}$ are only defined on two (overlapping) intervals of width $\pi$. As shown in Fig.~\ref{fig:Hadamard_test}(b), we introduce four distinct regions of width $\pi/2$, in which the phase $\theta$ has to be calculated based on different inversion formulas, defined as
\begin{alignat*}{2}
        &\text{Region 1:} \quad && \cos(\theta)< 0, \, \sin(\theta) < 0 \\
        & \quad && \rightarrow \theta =  -\pi + | \sin^{-1}(1-2P_{\text{sin}}(0))|, \\
        &\text{Region 2:} \quad && \cos(\theta)> 0, \, \sin(\theta) < 0 \\
        & \quad && \rightarrow \theta = \sin^{-1}(1-2P_{\text{sin}}(0)), \\
        &\text{Region 3:} \quad && \cos(\theta)> 0, \, \sin(\theta) < 0 \\
        & \quad && \rightarrow \theta = \cos^{-1}(1-2P_{\text{cos}}(0)), \\
        &\text{Region 4:} \quad && \cos(\theta)> 0, \, \sin(\theta) < 0 \\
        & \quad && \rightarrow \theta = \cos^{-1}(1-2P_{\text{cos}}(0)). \\
\end{alignat*}
Finally, we note that the operator $U$ is the time-evolution operator that evolves the eigenstate $\ket{n(\boldsymbol{k})}$ of the Hamiltonian $H(\boldsymbol{k})$ through a closed loop in momentum space. However, that doesn't affect how the Hadamard test scheme works, as the final state of the lower qubit line is equivalent to the initial state (up to a phase).

\section{Interacting systems}
\label{app:inter}

Here, we outline how the adiabatic time-evolution algorithm can be implemented for an interacting system. To this end, one needs to identify a suitable parameter in the Hamiltonian, whose evolution in parameter space gives rise to a topological invariant. For the sake of concreteness, we  consider the inter6acting Heisenberg spin-$1/2$ chain with frustration. This class of spin models describes a range of phenomena and requires a local topological classification to characterize their gapped quantum spin-liquid phase. The Hamiltonian reads
\begin{equation}
     H = \sum_{i, j} J_{i j} \boldsymbol{ 
 S}_i \cdot \boldsymbol{  S}_j,
\end{equation}
where the interaction strengths between spins $i$ and $j$ are denoted by $J_{i j}$. The following  analysis follows Ref.~\onlinecite{Hatsugai2007}, and we include it here for the sake of convenience.

First, we need to define a parameter, which gives rise to a topological invariant, thereby allowing us to identify a topological order. We may modify each link as 
\begin{equation}
    J_{ij} \boldsymbol{S}_i \cdot \boldsymbol{ S}_j \rightarrow J_{ij} \left(  \left( x^*  S_{i+}  S_{j-} + x S_{i-}  S_{j+} \right)/2 +  S_{iz} S_{jz} \right),
\end{equation}
where the local twist is described by the complex phase $x = e^{i \theta}$. This modification has the important property that an evolution $\theta = 0 \rightarrow  2\pi$ describes a closed loop in the parameter space spanned by $x$. In Ref.~\onlinecite{Hatsugai2007}, it was shown that by defining the Berry connection
\begin{equation}
    A_{\psi_0}(x) = \braket{\psi_0(x)}{\psi'_0(x)} 
\end{equation}
for the ground state $\psi_0(x)$, a well-defined and gauge-invariant Berry phase can be found as
\begin{equation}
    \gamma_C = - i \int_C dx A_{\psi_0}(x) = 0, \pi \, (\text{mod} \, 2\pi)
\end{equation}
for an evolution along a closed path $C$. Given these preparations, our algorithm can now be applied as described in our manuscript with the spins mapped onto the qubits of the quantum computer. In this case, the adiabatic evolution  incrementally increases the parameter $\theta$. The ground state will have to be prepared via a suitable routine, such as the variational quantum eigensolver. Finally, further decompositions of the adiabatic evolution via Trotterization may be necessary.

\bibliography{biblio}

%merlin.mbs apsrev4-1.bst 2010-07-25 4.21a (PWD, AO, DPC) hacked
%Control: key (0)
%Control: author (0) dotless jnrlst
%Control: editor formatted (1) identically to author
%Control: production of article title (0) allowed
%Control: page (1) range
%Control: year (0) verbatim
%Control: production of eprint (0) enabled
\begin{thebibliography}{61}%
\makeatletter
\providecommand \@ifxundefined [1]{%
 \@ifx{#1\undefined}
}%
\providecommand \@ifnum [1]{%
 \ifnum #1\expandafter \@firstoftwo
 \else \expandafter \@secondoftwo
 \fi
}%
\providecommand \@ifx [1]{%
 \ifx #1\expandafter \@firstoftwo
 \else \expandafter \@secondoftwo
 \fi
}%
\providecommand \natexlab [1]{#1}%
\providecommand \enquote  [1]{``#1''}%
\providecommand \bibnamefont  [1]{#1}%
\providecommand \bibfnamefont [1]{#1}%
\providecommand \citenamefont [1]{#1}%
\providecommand \href@noop [0]{\@secondoftwo}%
\providecommand \href [0]{\begingroup \@sanitize@url \@href}%
\providecommand \@href[1]{\@@startlink{#1}\@@href}%
\providecommand \@@href[1]{\endgroup#1\@@endlink}%
\providecommand \@sanitize@url [0]{\catcode `\\12\catcode `\$12\catcode
  `\&12\catcode `\#12\catcode `\^12\catcode `\_12\catcode `\%12\relax}%
\providecommand \@@startlink[1]{}%
\providecommand \@@endlink[0]{}%
\providecommand \url  [0]{\begingroup\@sanitize@url \@url }%
\providecommand \@url [1]{\endgroup\@href {#1}{\urlprefix }}%
\providecommand \urlprefix  [0]{URL }%
\providecommand \Eprint [0]{\href }%
\providecommand \doibase [0]{http://dx.doi.org/}%
\providecommand \selectlanguage [0]{\@gobble}%
\providecommand \bibinfo  [0]{\@secondoftwo}%
\providecommand \bibfield  [0]{\@secondoftwo}%
\providecommand \translation [1]{[#1]}%
\providecommand \BibitemOpen [0]{}%
\providecommand \bibitemStop [0]{}%
\providecommand \bibitemNoStop [0]{.\EOS\space}%
\providecommand \EOS [0]{\spacefactor3000\relax}%
\providecommand \BibitemShut  [1]{\csname bibitem#1\endcsname}%
\let\auto@bib@innerbib\@empty
%</preamble>
\bibitem [{\citenamefont {Qi}\ and\ \citenamefont
  {Zhang}(2011)}]{RevModPhys.83.1057}%
  \BibitemOpen
  \bibfield  {author} {\bibinfo {author} {\bibfnamefont {X.-L.}\ \bibnamefont
  {Qi}}\ and\ \bibinfo {author} {\bibfnamefont {S.-C.}\ \bibnamefont {Zhang}},\
  }\bibfield  {title} {\enquote {\bibinfo {title} {Topological insulators and
  superconductors},}\ }\href {\doibase 10.1103/RevModPhys.83.1057} {\bibfield
  {journal} {\bibinfo  {journal} {Rev. Mod. Phys.}\ }\textbf {\bibinfo {volume}
  {83}},\ \bibinfo {pages} {1057} (\bibinfo {year} {2011})}\BibitemShut
  {NoStop}%
\bibitem [{\citenamefont {Hasan}\ and\ \citenamefont
  {Kane}(2010)}]{RevModPhys.82.3045}%
  \BibitemOpen
  \bibfield  {author} {\bibinfo {author} {\bibfnamefont {M.~Z.}\ \bibnamefont
  {Hasan}}\ and\ \bibinfo {author} {\bibfnamefont {C.~L.}\ \bibnamefont
  {Kane}},\ }\bibfield  {title} {\enquote {\bibinfo {title} {Colloquium:
  Topological insulators},}\ }\href {\doibase 10.1103/RevModPhys.82.3045}
  {\bibfield  {journal} {\bibinfo  {journal} {Rev. Mod. Phys.}\ }\textbf
  {\bibinfo {volume} {82}},\ \bibinfo {pages} {3045} (\bibinfo {year}
  {2010})}\BibitemShut {NoStop}%
\bibitem [{\citenamefont {Moessner}\ and\ \citenamefont
  {Moore}(2021)}]{Moessner2021}%
  \BibitemOpen
  \bibfield  {author} {\bibinfo {author} {\bibfnamefont {R.}~\bibnamefont
  {Moessner}}\ and\ \bibinfo {author} {\bibfnamefont {J.~E.}\ \bibnamefont
  {Moore}},\ }\href {\doibase 10.1017/9781316226308} {\emph {\bibinfo {title}
  {Topological Phases of Matter}}}\ (\bibinfo  {publisher} {Cambridge
  University Press},\ \bibinfo {year} {2021})\BibitemShut {NoStop}%
\bibitem [{\citenamefont {J.~Asb{\'{o}}th}\ and\ \citenamefont
  {P{\'{a}}lyi}(2016)}]{asboth2016short}%
  \BibitemOpen
  \bibfield  {author} {\bibinfo {author} {\bibfnamefont {L.~Oroszl{\'{a}}ny}\
  \bibnamefont {J.~Asb{\'{o}}th}}\ and\ \bibinfo {author} {\bibfnamefont
  {A.}~\bibnamefont {P{\'{a}}lyi}},\ }\href {\doibase
  10.1007/978-3-319-25607-8} {\emph {\bibinfo {title} {A Short Course on
  Topological Insulators}}}\ (\bibinfo  {publisher} {Springer},\ \bibinfo
  {year} {2016})\BibitemShut {NoStop}%
\bibitem [{\citenamefont {Xiao}\ \emph {et~al.}(2010)\citenamefont {Xiao},
  \citenamefont {Chang},\ and\ \citenamefont
  {Niu}}]{Berry_phase_effects_elec_prop}%
  \BibitemOpen
  \bibfield  {author} {\bibinfo {author} {\bibfnamefont {D.}~\bibnamefont
  {Xiao}}, \bibinfo {author} {\bibfnamefont {M.-C.}\ \bibnamefont {Chang}}, \
  and\ \bibinfo {author} {\bibfnamefont {Q.}~\bibnamefont {Niu}},\ }\bibfield
  {title} {\enquote {\bibinfo {title} {Berry phase effects on electronic
  properties},}\ }\href {\doibase 10.1103/RevModPhys.82.1959} {\bibfield
  {journal} {\bibinfo  {journal} {Rev. Mod. Phys.}\ }\textbf {\bibinfo {volume}
  {82}},\ \bibinfo {pages} {1959} (\bibinfo {year} {2010})}\BibitemShut
  {NoStop}%
\bibitem [{\citenamefont {Ren}\ \emph {et~al.}(2016)\citenamefont {Ren},
  \citenamefont {Qiao},\ and\ \citenamefont {Niu}}]{Ren2016}%
  \BibitemOpen
  \bibfield  {author} {\bibinfo {author} {\bibfnamefont {Y.}~\bibnamefont
  {Ren}}, \bibinfo {author} {\bibfnamefont {Z.}~\bibnamefont {Qiao}}, \ and\
  \bibinfo {author} {\bibfnamefont {Q.}~\bibnamefont {Niu}},\ }\bibfield
  {title} {\enquote {\bibinfo {title} {Topological phases in two-dimensional
  materials: a review},}\ }\href {\doibase 10.1088/0034-4885/79/6/066501}
  {\bibfield  {journal} {\bibinfo  {journal} {Rep. Prog. Phys.}\ }\textbf
  {\bibinfo {volume} {79}},\ \bibinfo {pages} {066501} (\bibinfo {year}
  {2016})}\BibitemShut {NoStop}%
\bibitem [{\citenamefont {Chang}\ \emph {et~al.}(2023)\citenamefont {Chang},
  \citenamefont {Liu},\ and\ \citenamefont {MacDonald}}]{RevModPhys.95.011002}%
  \BibitemOpen
  \bibfield  {author} {\bibinfo {author} {\bibfnamefont {C.-Z.}\ \bibnamefont
  {Chang}}, \bibinfo {author} {\bibfnamefont {C.-X.}\ \bibnamefont {Liu}}, \
  and\ \bibinfo {author} {\bibfnamefont {A.~H.}\ \bibnamefont {MacDonald}},\
  }\bibfield  {title} {\enquote {\bibinfo {title} {{Colloquium: Quantum
  anomalous Hall effect}},}\ }\href {\doibase 10.1103/RevModPhys.95.011002}
  {\bibfield  {journal} {\bibinfo  {journal} {Rev. Mod. Phys.}\ }\textbf
  {\bibinfo {volume} {95}},\ \bibinfo {pages} {011002} (\bibinfo {year}
  {2023})}\BibitemShut {NoStop}%
\bibitem [{\citenamefont {Maciejko}\ \emph {et~al.}(2011)\citenamefont
  {Maciejko}, \citenamefont {Hughes},\ and\ \citenamefont
  {Zhang}}]{Maciejko2011}%
  \BibitemOpen
  \bibfield  {author} {\bibinfo {author} {\bibfnamefont {J.}~\bibnamefont
  {Maciejko}}, \bibinfo {author} {\bibfnamefont {T.~L.}\ \bibnamefont
  {Hughes}}, \ and\ \bibinfo {author} {\bibfnamefont {S.-C.}\ \bibnamefont
  {Zhang}},\ }\bibfield  {title} {\enquote {\bibinfo {title} {{The Quantum Spin
  Hall Effect}},}\ }\href {\doibase 10.1146/annurev-conmatphys-062910-140538}
  {\bibfield  {journal} {\bibinfo  {journal} {Annu. Rev. Condens. Matter
  Phys.}\ }\textbf {\bibinfo {volume} {2}},\ \bibinfo {pages} {31} (\bibinfo
  {year} {2011})}\BibitemShut {NoStop}%
\bibitem [{\citenamefont {Alicea}(2012)}]{Alicea2012}%
  \BibitemOpen
  \bibfield  {author} {\bibinfo {author} {\bibfnamefont {J.}~\bibnamefont
  {Alicea}},\ }\bibfield  {title} {\enquote {\bibinfo {title} {{New directions
  in the pursuit of Majorana fermions in solid state systems}},}\ }\href
  {\doibase 10.1088/0034-4885/75/7/076501} {\bibfield  {journal} {\bibinfo
  {journal} {Rep. Prog. Phys}\ }\textbf {\bibinfo {volume} {75}},\ \bibinfo
  {pages} {076501} (\bibinfo {year} {2012})}\BibitemShut {NoStop}%
\bibitem [{\citenamefont {Berry}(1984)}]{Berry_initial}%
  \BibitemOpen
  \bibfield  {author} {\bibinfo {author} {\bibfnamefont {M.~V.}\ \bibnamefont
  {Berry}},\ }\bibfield  {title} {\enquote {\bibinfo {title} {Quantal phase
  factors accompanying adiabatic changes},}\ }\href {\doibase
  10.1098/rspa.1984.0023} {\bibfield  {journal} {\bibinfo  {journal} {Proc. R.
  Soc. Lond. A}\ }\textbf {\bibinfo {volume} {392}},\ \bibinfo {pages} {45}
  (\bibinfo {year} {1984})}\BibitemShut {NoStop}%
\bibitem [{\citenamefont {Zak}(1989)}]{PRLZak}%
  \BibitemOpen
  \bibfield  {author} {\bibinfo {author} {\bibfnamefont {J.}~\bibnamefont
  {Zak}},\ }\bibfield  {title} {\enquote {\bibinfo {title} {Berry's phase for
  energy bands in solids},}\ }\href {\doibase 10.1103/PhysRevLett.62.2747}
  {\bibfield  {journal} {\bibinfo  {journal} {Phys. Rev. Lett.}\ }\textbf
  {\bibinfo {volume} {62}},\ \bibinfo {pages} {2747} (\bibinfo {year}
  {1989})}\BibitemShut {NoStop}%
\bibitem [{\citenamefont {Haldane}(1988)}]{Haldane1988}%
  \BibitemOpen
  \bibfield  {author} {\bibinfo {author} {\bibfnamefont {F.~D.~M.}\
  \bibnamefont {Haldane}},\ }\bibfield  {title} {\enquote {\bibinfo {title}
  {{Model for a Quantum Hall Effect without Landau Levels: Condensed-Matter
  Realization of the "Parity Anomaly"}},}\ }\href {\doibase
  10.1103/physrevlett.61.2015} {\bibfield  {journal} {\bibinfo  {journal}
  {Phys. Rev. Lett.}\ }\textbf {\bibinfo {volume} {61}},\ \bibinfo {pages}
  {2015} (\bibinfo {year} {1988})}\BibitemShut {NoStop}%
\bibitem [{\citenamefont {Kane}\ and\ \citenamefont
  {Mele}(2005)}]{quantum_spin_Hall_effect_Graphene}%
  \BibitemOpen
  \bibfield  {author} {\bibinfo {author} {\bibfnamefont {C.~L.}\ \bibnamefont
  {Kane}}\ and\ \bibinfo {author} {\bibfnamefont {E.~J.}\ \bibnamefont
  {Mele}},\ }\bibfield  {title} {\enquote {\bibinfo {title} {{Quantum Spin Hall
  Effect in Graphene}},}\ }\href {\doibase 10.1103/PhysRevLett.95.226801}
  {\bibfield  {journal} {\bibinfo  {journal} {Phys. Rev. Lett.}\ }\textbf
  {\bibinfo {volume} {95}},\ \bibinfo {pages} {226801} (\bibinfo {year}
  {2005})}\BibitemShut {NoStop}%
\bibitem [{\citenamefont {Kitaev}(2001)}]{Kitaev2001}%
  \BibitemOpen
  \bibfield  {author} {\bibinfo {author} {\bibfnamefont {A.~Yu.}\ \bibnamefont
  {Kitaev}},\ }\bibfield  {title} {\enquote {\bibinfo {title} {{Unpaired
  Majorana fermions in quantum wires}},}\ }\href {\doibase
  10.1070/1063-7869/44/10s/s29} {\bibfield  {journal} {\bibinfo  {journal}
  {Phys.-Usp.}\ }\textbf {\bibinfo {volume} {44}},\ \bibinfo {pages} {131}
  (\bibinfo {year} {2001})}\BibitemShut {NoStop}%
\bibitem [{\citenamefont {Alicea}\ \emph {et~al.}(2011)\citenamefont {Alicea},
  \citenamefont {Oreg}, \citenamefont {Refael}, \citenamefont {von Oppen},\
  and\ \citenamefont {Fisher}}]{Alicea2011}%
  \BibitemOpen
  \bibfield  {author} {\bibinfo {author} {\bibfnamefont {J.}~\bibnamefont
  {Alicea}}, \bibinfo {author} {\bibfnamefont {Y.}~\bibnamefont {Oreg}},
  \bibinfo {author} {\bibfnamefont {G.}~\bibnamefont {Refael}}, \bibinfo
  {author} {\bibfnamefont {F.}~\bibnamefont {von Oppen}}, \ and\ \bibinfo
  {author} {\bibfnamefont {M.~P.~A.}\ \bibnamefont {Fisher}},\ }\bibfield
  {title} {\enquote {\bibinfo {title} {{Non-Abelian statistics and topological
  quantum information processing in 1D wire networks}},}\ }\href {\doibase
  10.1038/nphys1915} {\bibfield  {journal} {\bibinfo  {journal} {Nat. Phys.}\
  }\textbf {\bibinfo {volume} {7}},\ \bibinfo {pages} {412} (\bibinfo {year}
  {2011})}\BibitemShut {NoStop}%
\bibitem [{\citenamefont {Nayak}\ \emph {et~al.}(2008)\citenamefont {Nayak},
  \citenamefont {Simon}, \citenamefont {Stern}, \citenamefont {Freedman},\ and\
  \citenamefont {Das~Sarma}}]{top_QC}%
  \BibitemOpen
  \bibfield  {author} {\bibinfo {author} {\bibfnamefont {C.}~\bibnamefont
  {Nayak}}, \bibinfo {author} {\bibfnamefont {S.~H.}\ \bibnamefont {Simon}},
  \bibinfo {author} {\bibfnamefont {A.}~\bibnamefont {Stern}}, \bibinfo
  {author} {\bibfnamefont {M.}~\bibnamefont {Freedman}}, \ and\ \bibinfo
  {author} {\bibfnamefont {S.}~\bibnamefont {Das~Sarma}},\ }\bibfield  {title}
  {\enquote {\bibinfo {title} {{Non-Abelian anyons and topological quantum
  computation}},}\ }\href {\doibase 10.1103/RevModPhys.80.1083} {\bibfield
  {journal} {\bibinfo  {journal} {Rev. Mod. Phys.}\ }\textbf {\bibinfo {volume}
  {80}},\ \bibinfo {pages} {1083} (\bibinfo {year} {2008})}\BibitemShut
  {NoStop}%
\bibitem [{\citenamefont {Lahtinen}\ and\ \citenamefont
  {Pachos}(2017)}]{Lahtinen2017}%
  \BibitemOpen
  \bibfield  {author} {\bibinfo {author} {\bibfnamefont {V.}~\bibnamefont
  {Lahtinen}}\ and\ \bibinfo {author} {\bibfnamefont {J.}~\bibnamefont
  {Pachos}},\ }\bibfield  {title} {\enquote {\bibinfo {title} {{A Short
  Introduction to Topological Quantum Computation}},}\ }\href
  {https://doi.org/10.21468/scipostphys.3.3.021} {\bibfield  {journal}
  {\bibinfo  {journal} {{SciPost} Phys.}\ }\textbf {\bibinfo {volume} {3}},\
  \bibinfo {pages} {21} (\bibinfo {year} {2017})}\BibitemShut {NoStop}%
\bibitem [{\citenamefont {T.~Fukui}\ and\ \citenamefont
  {Suzuki}(2005)}]{Fukui2005}%
  \BibitemOpen
  \bibfield  {author} {\bibinfo {author} {\bibfnamefont {Y.~Hatsugai}\
  \bibnamefont {T.~Fukui}}\ and\ \bibinfo {author} {\bibfnamefont
  {H.}~\bibnamefont {Suzuki}},\ }\bibfield  {title} {\enquote {\bibinfo {title}
  {{Chern Numbers in Discretized Brillouin Zone: Efficient Method of Computing
  (Spin) Hall Conductances}},}\ }\href {\doibase 10.1143/jpsj.74.1674}
  {\bibfield  {journal} {\bibinfo  {journal} {J. Phys. Soc. Jpn.}\ }\textbf
  {\bibinfo {volume} {74}},\ \bibinfo {pages} {1674} (\bibinfo {year}
  {2005})}\BibitemShut {NoStop}%
\bibitem [{\citenamefont {K.~Kudo}\ and\ \citenamefont
  {Hatsugai}(2019)}]{Chern_number_without_integration}%
  \BibitemOpen
  \bibfield  {author} {\bibinfo {author} {\bibfnamefont {T.~Kariyado}\
  \bibnamefont {K.~Kudo}, \bibfnamefont {H.~Watanabe}}\ and\ \bibinfo {author}
  {\bibfnamefont {Y.}~\bibnamefont {Hatsugai}},\ }\bibfield  {title} {\enquote
  {\bibinfo {title} {{Many-Body Chern Number without Integration}},}\ }\href
  {\doibase 10.1103/PhysRevLett.122.146601} {\bibfield  {journal} {\bibinfo
  {journal} {Phys. Rev. Lett.}\ }\textbf {\bibinfo {volume} {122}},\ \bibinfo
  {pages} {146601} (\bibinfo {year} {2019})}\BibitemShut {NoStop}%
\bibitem [{\citenamefont {{Ayral}}\ \emph {et~al.}(2023)\citenamefont
  {{Ayral}}, \citenamefont {{Besserve}}, \citenamefont {{Lacroix}},\ and\
  \citenamefont {{Ruiz Guzman}}}]{QC_many_body_physics}%
  \BibitemOpen
  \bibfield  {author} {\bibinfo {author} {\bibfnamefont {T.}~\bibnamefont
  {{Ayral}}}, \bibinfo {author} {\bibfnamefont {P.}~\bibnamefont {{Besserve}}},
  \bibinfo {author} {\bibfnamefont {D.}~\bibnamefont {{Lacroix}}}, \ and\
  \bibinfo {author} {\bibfnamefont {E.~A.}\ \bibnamefont {{Ruiz Guzman}}},\
  }\href@noop {} {\enquote {\bibinfo {title} {{Quantum computing with and for
  many-body physics}},}\ } (\bibinfo {year} {2023}),\ \Eprint
  {http://arxiv.org/abs/2303.04850} {arXiv:2303.04850 [quant-ph]} \BibitemShut
  {NoStop}%
\bibitem [{\citenamefont {B.~Murta}\ and\ \citenamefont
  {Fern\'andez-Rossier}(2020)}]{Berry_QC}%
  \BibitemOpen
  \bibfield  {author} {\bibinfo {author} {\bibfnamefont {G.~Catarina}\
  \bibnamefont {B.~Murta}}\ and\ \bibinfo {author} {\bibfnamefont
  {J.}~\bibnamefont {Fern\'andez-Rossier}},\ }\bibfield  {title} {\enquote
  {\bibinfo {title} {Berry phase estimation in gate-based adiabatic quantum
  simulation},}\ }\href {\doibase 10.1103/PhysRevA.101.020302} {\bibfield
  {journal} {\bibinfo  {journal} {Phys. Rev. A}\ }\textbf {\bibinfo {volume}
  {101}},\ \bibinfo {pages} {020302} (\bibinfo {year} {2020})}\BibitemShut
  {NoStop}%
\bibitem [{\citenamefont {Semeghini}\ \emph {et~al.}(2021)\citenamefont
  {Semeghini}, \citenamefont {Levine}, \citenamefont {Keesling}, \citenamefont
  {Ebadi}, \citenamefont {Wang}, \citenamefont {Bluvstein}, \citenamefont
  {Verresen}, \citenamefont {Pichler}, \citenamefont {Kalinowski},
  \citenamefont {Samajdar}, \citenamefont {Omran}, \citenamefont {Sachdev},
  \citenamefont {Vishwanath}, \citenamefont {Greiner}, \citenamefont
  {Vuleti{\'{c}}},\ and\ \citenamefont {Lukin}}]{Semeghini2021}%
  \BibitemOpen
  \bibfield  {author} {\bibinfo {author} {\bibfnamefont {G.}~\bibnamefont
  {Semeghini}}, \bibinfo {author} {\bibfnamefont {H.}~\bibnamefont {Levine}},
  \bibinfo {author} {\bibfnamefont {A.}~\bibnamefont {Keesling}}, \bibinfo
  {author} {\bibfnamefont {S.}~\bibnamefont {Ebadi}}, \bibinfo {author}
  {\bibfnamefont {T.~T.}\ \bibnamefont {Wang}}, \bibinfo {author}
  {\bibfnamefont {D.}~\bibnamefont {Bluvstein}}, \bibinfo {author}
  {\bibfnamefont {R.}~\bibnamefont {Verresen}}, \bibinfo {author}
  {\bibfnamefont {H.}~\bibnamefont {Pichler}}, \bibinfo {author} {\bibfnamefont
  {M.}~\bibnamefont {Kalinowski}}, \bibinfo {author} {\bibfnamefont
  {R.}~\bibnamefont {Samajdar}}, \bibinfo {author} {\bibfnamefont
  {A.}~\bibnamefont {Omran}}, \bibinfo {author} {\bibfnamefont
  {S.}~\bibnamefont {Sachdev}}, \bibinfo {author} {\bibfnamefont
  {A.}~\bibnamefont {Vishwanath}}, \bibinfo {author} {\bibfnamefont
  {M.}~\bibnamefont {Greiner}}, \bibinfo {author} {\bibfnamefont
  {V.}~\bibnamefont {Vuleti{\'{c}}}}, \ and\ \bibinfo {author} {\bibfnamefont
  {M.~D.}\ \bibnamefont {Lukin}},\ }\bibfield  {title} {\enquote {\bibinfo
  {title} {Probing topological spin liquids on a programmable quantum
  simulator},}\ }\href {\doibase 10.1126/science.abi8794} {\bibfield  {journal}
  {\bibinfo  {journal} {Science}\ }\textbf {\bibinfo {volume} {374}},\ \bibinfo
  {pages} {1242} (\bibinfo {year} {2021})}\BibitemShut {NoStop}%
\bibitem [{\citenamefont {X.~Xiao}\ and\ \citenamefont
  {Kemper}(2021)}]{Xiao2021determiningquantum}%
  \BibitemOpen
  \bibfield  {author} {\bibinfo {author} {\bibfnamefont {J.~K.~Freericks}\
  \bibnamefont {X.~Xiao}}\ and\ \bibinfo {author} {\bibfnamefont {A.~F.}\
  \bibnamefont {Kemper}},\ }\bibfield  {title} {\enquote {\bibinfo {title}
  {Determining quantum phase diagrams of topological {K}itaev-inspired models
  on {NISQ} quantum hardware},}\ }\href {\doibase 10.22331/q-2021-09-28-553}
  {\bibfield  {journal} {\bibinfo  {journal} {{Quantum}}\ }\textbf {\bibinfo
  {volume} {5}},\ \bibinfo {pages} {553} (\bibinfo {year} {2021})}\BibitemShut
  {NoStop}%
\bibitem [{\citenamefont {X.~Xiao}\ and\ \citenamefont
  {Kemper}(2023)}]{Xiao2023robustmeasurementof}%
  \BibitemOpen
  \bibfield  {author} {\bibinfo {author} {\bibfnamefont {J.~K.~Freericks}\
  \bibnamefont {X.~Xiao}}\ and\ \bibinfo {author} {\bibfnamefont {A.~F.}\
  \bibnamefont {Kemper}},\ }\bibfield  {title} {\enquote {\bibinfo {title}
  {Robust measurement of wave function topology on {NISQ} quantum computers},}\
  }\href {\doibase 10.22331/q-2023-04-27-987} {\bibfield  {journal} {\bibinfo
  {journal} {{Quantum}}\ }\textbf {\bibinfo {volume} {7}},\ \bibinfo {pages}
  {987} (\bibinfo {year} {2023})}\BibitemShut {NoStop}%
\bibitem [{\citenamefont {{Sugimoto}}(2023)}]{QC_alg_many_body_top_invariants}%
  \BibitemOpen
  \bibfield  {author} {\bibinfo {author} {\bibfnamefont {T.}~\bibnamefont
  {{Sugimoto}}},\ }\href@noop {} {\enquote {\bibinfo {title} {{Quantum-circuit
  algorithms for many-body topological invariant and Majorana zero mode}},}\ }
  (\bibinfo {year} {2023}),\ \Eprint {http://arxiv.org/abs/2304.13408}
  {arXiv:2304.13408} \BibitemShut {NoStop}%
\bibitem [{\citenamefont {Mei}\ \emph {et~al.}(2020)\citenamefont {Mei},
  \citenamefont {Guo}, \citenamefont {Yu}, \citenamefont {Xiao}, \citenamefont
  {Zhu},\ and\ \citenamefont
  {Jia}}]{dig_simulation_top_matter_programmable_quantum_processors}%
  \BibitemOpen
  \bibfield  {author} {\bibinfo {author} {\bibfnamefont {F.}~\bibnamefont
  {Mei}}, \bibinfo {author} {\bibfnamefont {Q.}~\bibnamefont {Guo}}, \bibinfo
  {author} {\bibfnamefont {Y.-F.}\ \bibnamefont {Yu}}, \bibinfo {author}
  {\bibfnamefont {L.}~\bibnamefont {Xiao}}, \bibinfo {author} {\bibfnamefont
  {S.-L.}\ \bibnamefont {Zhu}}, \ and\ \bibinfo {author} {\bibfnamefont
  {S.}~\bibnamefont {Jia}},\ }\bibfield  {title} {\enquote {\bibinfo {title}
  {{Digital Simulation of Topological Matter on Programmable Quantum
  Processors}},}\ }\href {\doibase 10.1103/PhysRevLett.125.160503} {\bibfield
  {journal} {\bibinfo  {journal} {Phys. Rev. Lett.}\ }\textbf {\bibinfo
  {volume} {125}},\ \bibinfo {pages} {160503} (\bibinfo {year}
  {2020})}\BibitemShut {NoStop}%
\bibitem [{\citenamefont {Smith}\ \emph {et~al.}(2022)\citenamefont {Smith},
  \citenamefont {Jobst}, \citenamefont {Green},\ and\ \citenamefont
  {Pollmann}}]{crossing_top_phase_transition_QC}%
  \BibitemOpen
  \bibfield  {author} {\bibinfo {author} {\bibfnamefont {A.}~\bibnamefont
  {Smith}}, \bibinfo {author} {\bibfnamefont {B.}~\bibnamefont {Jobst}},
  \bibinfo {author} {\bibfnamefont {A.~G.}\ \bibnamefont {Green}}, \ and\
  \bibinfo {author} {\bibfnamefont {F.}~\bibnamefont {Pollmann}},\ }\bibfield
  {title} {\enquote {\bibinfo {title} {Crossing a topological phase transition
  with a quantum computer},}\ }\href {\doibase
  10.1103/PhysRevResearch.4.L022020} {\bibfield  {journal} {\bibinfo  {journal}
  {Phys. Rev. Res.}\ }\textbf {\bibinfo {volume} {4}},\ \bibinfo {pages}
  {L022020} (\bibinfo {year} {2022})}\BibitemShut {NoStop}%
\bibitem [{\citenamefont {Azses}\ \emph {et~al.}(2020)\citenamefont {Azses},
  \citenamefont {Haenel}, \citenamefont {Naveh}, \citenamefont {Raussendorf},
  \citenamefont {Sela},\ and\ \citenamefont
  {Torre}}]{identification_sym_protected_top_states_QC}%
  \BibitemOpen
  \bibfield  {author} {\bibinfo {author} {\bibfnamefont {D.}~\bibnamefont
  {Azses}}, \bibinfo {author} {\bibfnamefont {R.}~\bibnamefont {Haenel}},
  \bibinfo {author} {\bibfnamefont {Y.}~\bibnamefont {Naveh}}, \bibinfo
  {author} {\bibfnamefont {R.}~\bibnamefont {Raussendorf}}, \bibinfo {author}
  {\bibfnamefont {E.}~\bibnamefont {Sela}}, \ and\ \bibinfo {author}
  {\bibfnamefont {E.~G.~Dalla}\ \bibnamefont {Torre}},\ }\bibfield  {title}
  {\enquote {\bibinfo {title} {{Identification of Symmetry-Protected
  Topological States on Noisy Quantum Computers}},}\ }\href {\doibase
  10.1103/PhysRevLett.125.120502} {\bibfield  {journal} {\bibinfo  {journal}
  {Phys. Rev. Lett.}\ }\textbf {\bibinfo {volume} {125}},\ \bibinfo {pages}
  {120502} (\bibinfo {year} {2020})}\BibitemShut {NoStop}%
\bibitem [{\citenamefont {K.~Choo}\ and\ \citenamefont
  {Neupert}(2018)}]{measurement_ent_spectrum_sym_prot_top_state}%
  \BibitemOpen
  \bibfield  {author} {\bibinfo {author} {\bibfnamefont {N.~Regnault}\
  \bibnamefont {K.~Choo}, \bibfnamefont {C.~W. von~Keyserlingk}}\ and\ \bibinfo
  {author} {\bibfnamefont {T.}~\bibnamefont {Neupert}},\ }\bibfield  {title}
  {\enquote {\bibinfo {title} {{Measurement of the Entanglement Spectrum of a
  Symmetry-Protected Topological State Using the IBM Quantum Computer}},}\
  }\href {\doibase 10.1103/PhysRevLett.121.086808} {\bibfield  {journal}
  {\bibinfo  {journal} {Phys. Rev. Lett.}\ }\textbf {\bibinfo {volume} {121}},\
  \bibinfo {pages} {086808} (\bibinfo {year} {2018})}\BibitemShut {NoStop}%
\bibitem [{\citenamefont {Roushan}\ \emph {et~al.}(2014)\citenamefont
  {Roushan}, \citenamefont {Neill}, \citenamefont {Chen}, \citenamefont
  {Kolodrubetz}, \citenamefont {Quintana}, \citenamefont {Leung}, \citenamefont
  {Fang}, \citenamefont {Barends}, \citenamefont {Campbell}, \citenamefont
  {Chen}, \citenamefont {Chiaro}, \citenamefont {Dunsworth}, \citenamefont
  {Jeffrey}, \citenamefont {Kelly}, \citenamefont {Megrant}, \citenamefont
  {Mutus}, \citenamefont {O'Malley}, \citenamefont {Sank}, \citenamefont
  {Vainsencher}, \citenamefont {Wenner}, \citenamefont {White}, \citenamefont
  {Polkovnikov}, \citenamefont {Cleland},\ and\ \citenamefont
  {Martinis}}]{Roushan2014}%
  \BibitemOpen
  \bibfield  {author} {\bibinfo {author} {\bibfnamefont {P.}~\bibnamefont
  {Roushan}}, \bibinfo {author} {\bibfnamefont {C.}~\bibnamefont {Neill}},
  \bibinfo {author} {\bibfnamefont {Yu}~\bibnamefont {Chen}}, \bibinfo {author}
  {\bibfnamefont {M.}~\bibnamefont {Kolodrubetz}}, \bibinfo {author}
  {\bibfnamefont {C.}~\bibnamefont {Quintana}}, \bibinfo {author}
  {\bibfnamefont {N.}~\bibnamefont {Leung}}, \bibinfo {author} {\bibfnamefont
  {M.}~\bibnamefont {Fang}}, \bibinfo {author} {\bibfnamefont {R.}~\bibnamefont
  {Barends}}, \bibinfo {author} {\bibfnamefont {B.}~\bibnamefont {Campbell}},
  \bibinfo {author} {\bibfnamefont {Z.}~\bibnamefont {Chen}}, \bibinfo {author}
  {\bibfnamefont {B.}~\bibnamefont {Chiaro}}, \bibinfo {author} {\bibfnamefont
  {A.}~\bibnamefont {Dunsworth}}, \bibinfo {author} {\bibfnamefont
  {E.}~\bibnamefont {Jeffrey}}, \bibinfo {author} {\bibfnamefont
  {J.}~\bibnamefont {Kelly}}, \bibinfo {author} {\bibfnamefont
  {A.}~\bibnamefont {Megrant}}, \bibinfo {author} {\bibfnamefont
  {J.}~\bibnamefont {Mutus}}, \bibinfo {author} {\bibfnamefont {P.~J.~J.}\
  \bibnamefont {O'Malley}}, \bibinfo {author} {\bibfnamefont {D.}~\bibnamefont
  {Sank}}, \bibinfo {author} {\bibfnamefont {A.}~\bibnamefont {Vainsencher}},
  \bibinfo {author} {\bibfnamefont {J.}~\bibnamefont {Wenner}}, \bibinfo
  {author} {\bibfnamefont {T.}~\bibnamefont {White}}, \bibinfo {author}
  {\bibfnamefont {A.}~\bibnamefont {Polkovnikov}}, \bibinfo {author}
  {\bibfnamefont {A.~N.}\ \bibnamefont {Cleland}}, \ and\ \bibinfo {author}
  {\bibfnamefont {J.~M.}\ \bibnamefont {Martinis}},\ }\bibfield  {title}
  {\enquote {\bibinfo {title} {Observation of topological transitions in
  interacting quantum circuits},}\ }\href {\doibase 10.1038/nature13891}
  {\bibfield  {journal} {\bibinfo  {journal} {Nature}\ }\textbf {\bibinfo
  {volume} {515}},\ \bibinfo {pages} {241} (\bibinfo {year}
  {2014})}\BibitemShut {NoStop}%
\bibitem [{\citenamefont {Flurin}\ \emph {et~al.}(2017)\citenamefont {Flurin},
  \citenamefont {Ramasesh}, \citenamefont {Hacohen-Gourgy}, \citenamefont
  {Martin}, \citenamefont {Yao},\ and\ \citenamefont
  {Siddiqi}}]{observing_top_invariants_quantum_walks}%
  \BibitemOpen
  \bibfield  {author} {\bibinfo {author} {\bibfnamefont {E.}~\bibnamefont
  {Flurin}}, \bibinfo {author} {\bibfnamefont {V.~V.}\ \bibnamefont
  {Ramasesh}}, \bibinfo {author} {\bibfnamefont {S.}~\bibnamefont
  {Hacohen-Gourgy}}, \bibinfo {author} {\bibfnamefont {L.~S.}\ \bibnamefont
  {Martin}}, \bibinfo {author} {\bibfnamefont {N.~Y.}\ \bibnamefont {Yao}}, \
  and\ \bibinfo {author} {\bibfnamefont {I.}~\bibnamefont {Siddiqi}},\
  }\bibfield  {title} {\enquote {\bibinfo {title} {{Observing Topological
  Invariants Using Quantum Walks in Superconducting Circuits}},}\ }\href
  {\doibase 10.1103/PhysRevX.7.031023} {\bibfield  {journal} {\bibinfo
  {journal} {Phys. Rev. X}\ }\textbf {\bibinfo {volume} {7}},\ \bibinfo {pages}
  {031023} (\bibinfo {year} {2017})}\BibitemShut {NoStop}%
\bibitem [{\citenamefont {Xu}\ \emph {et~al.}(2018)\citenamefont {Xu},
  \citenamefont {Wang}, \citenamefont {Pan}, \citenamefont {Sun}, \citenamefont
  {Xu}, \citenamefont {Chen}, \citenamefont {Tang}, \citenamefont {Gong},
  \citenamefont {Han}, \citenamefont {Li},\ and\ \citenamefont
  {Guo}}]{winding_number_large_scale_chiral_quantum_walk}%
  \BibitemOpen
  \bibfield  {author} {\bibinfo {author} {\bibfnamefont {X.-Y.}\ \bibnamefont
  {Xu}}, \bibinfo {author} {\bibfnamefont {Q.-Q.}\ \bibnamefont {Wang}},
  \bibinfo {author} {\bibfnamefont {W.-W.}\ \bibnamefont {Pan}}, \bibinfo
  {author} {\bibfnamefont {K.}~\bibnamefont {Sun}}, \bibinfo {author}
  {\bibfnamefont {J.-S.}\ \bibnamefont {Xu}}, \bibinfo {author} {\bibfnamefont
  {G.}~\bibnamefont {Chen}}, \bibinfo {author} {\bibfnamefont {J.-S.}\
  \bibnamefont {Tang}}, \bibinfo {author} {\bibfnamefont {M.}~\bibnamefont
  {Gong}}, \bibinfo {author} {\bibfnamefont {Y.-J.}\ \bibnamefont {Han}},
  \bibinfo {author} {\bibfnamefont {C.-F.}\ \bibnamefont {Li}}, \ and\ \bibinfo
  {author} {\bibfnamefont {G.-C.}\ \bibnamefont {Guo}},\ }\bibfield  {title}
  {\enquote {\bibinfo {title} {{Measuring the Winding Number in a Large-Scale
  Chiral Quantum Walk}},}\ }\href {\doibase 10.1103/PhysRevLett.120.260501}
  {\bibfield  {journal} {\bibinfo  {journal} {Phys. Rev. Lett.}\ }\textbf
  {\bibinfo {volume} {120}},\ \bibinfo {pages} {260501} (\bibinfo {year}
  {2018})}\BibitemShut {NoStop}%
\bibitem [{\citenamefont {Zhan}\ \emph {et~al.}(2017)\citenamefont {Zhan},
  \citenamefont {Xiao}, \citenamefont {Bian}, \citenamefont {Wang},
  \citenamefont {Qiu}, \citenamefont {Sanders}, \citenamefont {Yi},\ and\
  \citenamefont {Xue}}]{top_inv_non_unitary_discrete_time_quantum_walks}%
  \BibitemOpen
  \bibfield  {author} {\bibinfo {author} {\bibfnamefont {X.}~\bibnamefont
  {Zhan}}, \bibinfo {author} {\bibfnamefont {L.}~\bibnamefont {Xiao}}, \bibinfo
  {author} {\bibfnamefont {Z.}~\bibnamefont {Bian}}, \bibinfo {author}
  {\bibfnamefont {K.}~\bibnamefont {Wang}}, \bibinfo {author} {\bibfnamefont
  {X.}~\bibnamefont {Qiu}}, \bibinfo {author} {\bibfnamefont {B.~C.}\
  \bibnamefont {Sanders}}, \bibinfo {author} {\bibfnamefont {W.}~\bibnamefont
  {Yi}}, \ and\ \bibinfo {author} {\bibfnamefont {P.}~\bibnamefont {Xue}},\
  }\bibfield  {title} {\enquote {\bibinfo {title} {{Detecting Topological
  Invariants in Nonunitary Discrete-Time Quantum Walks}},}\ }\href {\doibase
  10.1103/PhysRevLett.119.130501} {\bibfield  {journal} {\bibinfo  {journal}
  {Phys. Rev. Lett.}\ }\textbf {\bibinfo {volume} {119}},\ \bibinfo {pages}
  {130501} (\bibinfo {year} {2017})}\BibitemShut {NoStop}%
\bibitem [{\citenamefont {R.-Y.~{Sun}}\ and\ \citenamefont
  {{Yunoki}}(2023)}]{var_QC_corr_top_phases}%
  \BibitemOpen
  \bibfield  {author} {\bibinfo {author} {\bibfnamefont {T.~{Shirakawa}}\
  \bibnamefont {R.-Y.~{Sun}}}\ and\ \bibinfo {author} {\bibfnamefont
  {S.}~\bibnamefont {{Yunoki}}},\ }\href@noop {} {\enquote {\bibinfo {title}
  {{Efficient variational quantum circuit structure for correlated topological
  phases}},}\ } (\bibinfo {year} {2023}),\ \Eprint
  {http://arxiv.org/abs/2303.17187} {arXiv:2303.17187} \BibitemShut {NoStop}%
\bibitem [{\citenamefont {Tamiya}\ \emph {et~al.}(2021)\citenamefont {Tamiya},
  \citenamefont {Koh},\ and\ \citenamefont
  {Nakagawa}}]{PhysRevResearch.3.023244}%
  \BibitemOpen
  \bibfield  {author} {\bibinfo {author} {\bibfnamefont {Shiro}\ \bibnamefont
  {Tamiya}}, \bibinfo {author} {\bibfnamefont {Sho}\ \bibnamefont {Koh}}, \
  and\ \bibinfo {author} {\bibfnamefont {Yuya~O.}\ \bibnamefont {Nakagawa}},\
  }\bibfield  {title} {\enquote {\bibinfo {title} {Calculating nonadiabatic
  couplings and berry's phase by variational quantum eigensolvers},}\ }\href
  {\doibase 10.1103/PhysRevResearch.3.023244} {\bibfield  {journal} {\bibinfo
  {journal} {Phys. Rev. Res.}\ }\textbf {\bibinfo {volume} {3}},\ \bibinfo
  {pages} {023244} (\bibinfo {year} {2021})}\BibitemShut {NoStop}%
\bibitem [{Hel(2024{\natexlab{a}})}]{Helmi_technical}%
  \BibitemOpen
  \href {https://docs.csc.fi/computing/quantum-computing/helmi/helmi-specs/}
  {\enquote {\bibinfo {title} {Technical details about helmi},}\ } (\bibinfo
  {year} {2024}{\natexlab{a}}),\ \bibinfo {note} {accessed:
  2024-03-06}\BibitemShut {NoStop}%
\bibitem [{\citenamefont {{Luongo}}(2023)}]{Hadamard_test}%
  \BibitemOpen
  \bibfield  {author} {\bibinfo {author} {\bibfnamefont {A.}~\bibnamefont
  {{Luongo}}},\ }\href
  {https://quantumalgorithms.org/chapter-intro.html#hadamard-test} {\enquote
  {\bibinfo {title} {Quantum algorithms for data analysis},}\ } (\bibinfo
  {year} {2023}),\ \bibinfo {note} {accessed: 2023-10-12}\BibitemShut {NoStop}%
\bibitem [{\citenamefont {{Kitaev}}(1995)}]{Kitaev_QPE}%
  \BibitemOpen
  \bibfield  {author} {\bibinfo {author} {\bibfnamefont {A.~Yu.}\ \bibnamefont
  {{Kitaev}}},\ }\href@noop {} {\enquote {\bibinfo {title} {{Quantum
  measurements and the Abelian Stabilizer Problem}},}\ } (\bibinfo {year}
  {1995}),\ \Eprint {http://arxiv.org/abs/quant-ph/9511026}
  {arXiv:quant-ph/9511026} \BibitemShut {NoStop}%
\bibitem [{\citenamefont {Nielsen}\ and\ \citenamefont
  {Chuang}(2010)}]{nielsen_chuang_2010}%
  \BibitemOpen
  \bibfield  {author} {\bibinfo {author} {\bibfnamefont {M.~A.}\ \bibnamefont
  {Nielsen}}\ and\ \bibinfo {author} {\bibfnamefont {I.~L.}\ \bibnamefont
  {Chuang}},\ }\href {\doibase 10.1017/CBO9780511976667} {\emph {\bibinfo
  {title} {Quantum Computation and Quantum Information: 10th Anniversary
  Edition}}}\ (\bibinfo  {publisher} {Cambridge University Press},\ \bibinfo
  {year} {2010})\BibitemShut {NoStop}%
\bibitem [{\citenamefont {Marzari}\ and\ \citenamefont
  {Vanderbilt}(1997)}]{PhysRevB.56.12847}%
  \BibitemOpen
  \bibfield  {author} {\bibinfo {author} {\bibfnamefont {N.}~\bibnamefont
  {Marzari}}\ and\ \bibinfo {author} {\bibfnamefont {D.}~\bibnamefont
  {Vanderbilt}},\ }\bibfield  {title} {\enquote {\bibinfo {title} {{Maximally
  localized generalized Wannier functions for composite energy bands}},}\
  }\href {\doibase 10.1103/PhysRevB.56.12847} {\bibfield  {journal} {\bibinfo
  {journal} {Phys. Rev. B}\ }\textbf {\bibinfo {volume} {56}},\ \bibinfo
  {pages} {12847} (\bibinfo {year} {1997})}\BibitemShut {NoStop}%
\bibitem [{\citenamefont {Coh}\ and\ \citenamefont
  {Vanderbilt}(2009)}]{PhysRevLett.102.107603}%
  \BibitemOpen
  \bibfield  {author} {\bibinfo {author} {\bibfnamefont {S.}~\bibnamefont
  {Coh}}\ and\ \bibinfo {author} {\bibfnamefont {D.}~\bibnamefont
  {Vanderbilt}},\ }\bibfield  {title} {\enquote {\bibinfo {title} {{Electric
  Polarization in a Chern Insulator}},}\ }\href {\doibase
  10.1103/PhysRevLett.102.107603} {\bibfield  {journal} {\bibinfo  {journal}
  {Phys. Rev. Lett.}\ }\textbf {\bibinfo {volume} {102}},\ \bibinfo {pages}
  {107603} (\bibinfo {year} {2009})}\BibitemShut {NoStop}%
\bibitem [{\citenamefont {Wu}\ \emph {et~al.}(2018)\citenamefont {Wu},
  \citenamefont {Zhang}, \citenamefont {Song}, \citenamefont {Troyer},\ and\
  \citenamefont {Soluyanov}}]{Wu2018}%
  \BibitemOpen
  \bibfield  {author} {\bibinfo {author} {\bibfnamefont {Q.}~\bibnamefont
  {Wu}}, \bibinfo {author} {\bibfnamefont {S.}~\bibnamefont {Zhang}}, \bibinfo
  {author} {\bibfnamefont {H.-F.}\ \bibnamefont {Song}}, \bibinfo {author}
  {\bibfnamefont {M.}~\bibnamefont {Troyer}}, \ and\ \bibinfo {author}
  {\bibfnamefont {A.~A.}\ \bibnamefont {Soluyanov}},\ }\bibfield  {title}
  {\enquote {\bibinfo {title} {Wanniertools: An open-source software package
  for novel topological materials},}\ }\href {\doibase
  10.1016/j.cpc.2017.09.033} {\bibfield  {journal} {\bibinfo  {journal}
  {Comput. Phys. Commun.}\ }\textbf {\bibinfo {volume} {224}},\ \bibinfo
  {pages} {405} (\bibinfo {year} {2018})}\BibitemShut {NoStop}%
\bibitem [{\citenamefont {Yu}\ \emph {et~al.}(2011)\citenamefont {Yu},
  \citenamefont {Qi}, \citenamefont {Bernevig}, \citenamefont {Fang},\ and\
  \citenamefont {Dai}}]{PhysRevB.84.075119}%
  \BibitemOpen
  \bibfield  {author} {\bibinfo {author} {\bibfnamefont {R.}~\bibnamefont
  {Yu}}, \bibinfo {author} {\bibfnamefont {X.~L.}\ \bibnamefont {Qi}}, \bibinfo
  {author} {\bibfnamefont {A.}~\bibnamefont {Bernevig}}, \bibinfo {author}
  {\bibfnamefont {Z.}~\bibnamefont {Fang}}, \ and\ \bibinfo {author}
  {\bibfnamefont {X.}~\bibnamefont {Dai}},\ }\bibfield  {title} {\enquote
  {\bibinfo {title} {{Equivalent expression of ${\mathbb{Z}}_{2}$ topological
  invariant for band insulators using the non-Abelian Berry connection}},}\
  }\href {\doibase 10.1103/PhysRevB.84.075119} {\bibfield  {journal} {\bibinfo
  {journal} {Phys. Rev. B}\ }\textbf {\bibinfo {volume} {84}},\ \bibinfo
  {pages} {075119} (\bibinfo {year} {2011})}\BibitemShut {NoStop}%
\bibitem [{\citenamefont {Gresch}\ \emph {et~al.}(2017)\citenamefont {Gresch},
  \citenamefont {Aut\`es}, \citenamefont {Yazyev}, \citenamefont {Troyer},
  \citenamefont {Vanderbilt}, \citenamefont {Bernevig},\ and\ \citenamefont
  {Soluyanov}}]{PhysRevB.95.075146}%
  \BibitemOpen
  \bibfield  {author} {\bibinfo {author} {\bibfnamefont {D.}~\bibnamefont
  {Gresch}}, \bibinfo {author} {\bibfnamefont {G.}~\bibnamefont {Aut\`es}},
  \bibinfo {author} {\bibfnamefont {O.~V.}\ \bibnamefont {Yazyev}}, \bibinfo
  {author} {\bibfnamefont {M.}~\bibnamefont {Troyer}}, \bibinfo {author}
  {\bibfnamefont {D.}~\bibnamefont {Vanderbilt}}, \bibinfo {author}
  {\bibfnamefont {B.~A.}\ \bibnamefont {Bernevig}}, \ and\ \bibinfo {author}
  {\bibfnamefont {A.~A.}\ \bibnamefont {Soluyanov}},\ }\bibfield  {title}
  {\enquote {\bibinfo {title} {{Z2Pack: Numerical implementation of hybrid
  Wannier centers for identifying topological materials}},}\ }\href {\doibase
  10.1103/PhysRevB.95.075146} {\bibfield  {journal} {\bibinfo  {journal} {Phys.
  Rev. B}\ }\textbf {\bibinfo {volume} {95}},\ \bibinfo {pages} {075146}
  (\bibinfo {year} {2017})}\BibitemShut {NoStop}%
\bibitem [{\citenamefont {Soluyanov}\ and\ \citenamefont
  {Vanderbilt}(2011)}]{PhysRevB.83.235401}%
  \BibitemOpen
  \bibfield  {author} {\bibinfo {author} {\bibfnamefont {A.~A.}\ \bibnamefont
  {Soluyanov}}\ and\ \bibinfo {author} {\bibfnamefont {D.}~\bibnamefont
  {Vanderbilt}},\ }\bibfield  {title} {\enquote {\bibinfo {title} {Computing
  topological invariants without inversion symmetry},}\ }\href {\doibase
  10.1103/PhysRevB.83.235401} {\bibfield  {journal} {\bibinfo  {journal} {Phys.
  Rev. B}\ }\textbf {\bibinfo {volume} {83}},\ \bibinfo {pages} {235401}
  (\bibinfo {year} {2011})}\BibitemShut {NoStop}%
\bibitem [{\citenamefont {Tilly}\ \emph {et~al.}(2022)\citenamefont {Tilly},
  \citenamefont {Chen}, \citenamefont {Cao}, \citenamefont {Picozzi},
  \citenamefont {Setia}, \citenamefont {Li}, \citenamefont {Grant},
  \citenamefont {Wossnig}, \citenamefont {Rungger}, \citenamefont {Booth},\
  and\ \citenamefont {Tennyson}}]{Tilly2022}%
  \BibitemOpen
  \bibfield  {author} {\bibinfo {author} {\bibfnamefont {J.}~\bibnamefont
  {Tilly}}, \bibinfo {author} {\bibfnamefont {H.}~\bibnamefont {Chen}},
  \bibinfo {author} {\bibfnamefont {S.}~\bibnamefont {Cao}}, \bibinfo {author}
  {\bibfnamefont {D.}~\bibnamefont {Picozzi}}, \bibinfo {author} {\bibfnamefont
  {K.}~\bibnamefont {Setia}}, \bibinfo {author} {\bibfnamefont
  {Y.}~\bibnamefont {Li}}, \bibinfo {author} {\bibfnamefont {E.}~\bibnamefont
  {Grant}}, \bibinfo {author} {\bibfnamefont {L.}~\bibnamefont {Wossnig}},
  \bibinfo {author} {\bibfnamefont {I.}~\bibnamefont {Rungger}}, \bibinfo
  {author} {\bibfnamefont {H.~H.}\ \bibnamefont {Booth}}, \ and\ \bibinfo
  {author} {\bibfnamefont {J.}~\bibnamefont {Tennyson}},\ }\bibfield  {title}
  {\enquote {\bibinfo {title} {The variational quantum eigensolver: A review of
  methods and best practices},}\ }\href {\doibase
  10.1016/j.physrep.2022.08.003} {\bibfield  {journal} {\bibinfo  {journal}
  {Phys. Rep.}\ }\textbf {\bibinfo {volume} {986}},\ \bibinfo {pages} {1}
  (\bibinfo {year} {2022})}\BibitemShut {NoStop}%
\bibitem [{Note1()}]{Note1}%
  \BibitemOpen
  \bibinfo {note} {All of these values are subject to daily fluctuations and
  are averaged over all five qubits~\cite {Helmi_VTT}}\BibitemShut {NoStop}%
\bibitem [{\citenamefont {Schollwöck}(2011)}]{Schollwock_DMRG}%
  \BibitemOpen
  \bibfield  {author} {\bibinfo {author} {\bibfnamefont {U.}~\bibnamefont
  {Schollwöck}},\ }\bibfield  {title} {\enquote {\bibinfo {title} {The
  density-matrix renormalization group in the age of matrix product states},}\
  }\href {\doibase https://doi.org/10.1016/j.aop.2010.09.012} {\bibfield
  {journal} {\bibinfo  {journal} {Ann. Phys.}\ }\textbf {\bibinfo {volume}
  {326}},\ \bibinfo {pages} {96} (\bibinfo {year} {2011})}\BibitemShut
  {NoStop}%
\bibitem [{\citenamefont {{Niedermeier}}\ \emph {et~al.}(2023)\citenamefont
  {{Niedermeier}}, \citenamefont {{Lado}},\ and\ \citenamefont
  {{Flindt}}}]{TN_quantum_sim_lim_ent}%
  \BibitemOpen
  \bibfield  {author} {\bibinfo {author} {\bibfnamefont {M.}~\bibnamefont
  {{Niedermeier}}}, \bibinfo {author} {\bibfnamefont {J.~L.}\ \bibnamefont
  {{Lado}}}, \ and\ \bibinfo {author} {\bibfnamefont {C.}~\bibnamefont
  {{Flindt}}},\ }\href@noop {} {\enquote {\bibinfo {title} {{Tensor-Network
  Simulations of Noisy Quantum Computers}},}\ } (\bibinfo {year} {2023}),\
  \Eprint {http://arxiv.org/abs/2304.01751} {arXiv:2304.01751} \BibitemShut
  {NoStop}%
\bibitem [{\citenamefont {Zhou}\ \emph {et~al.}(2020)\citenamefont {Zhou},
  \citenamefont {Stoudenmire},\ and\ \citenamefont
  {Waintal}}]{PRX_what_limits}%
  \BibitemOpen
  \bibfield  {author} {\bibinfo {author} {\bibfnamefont {Y.}~\bibnamefont
  {Zhou}}, \bibinfo {author} {\bibfnamefont {E.~M.}\ \bibnamefont
  {Stoudenmire}}, \ and\ \bibinfo {author} {\bibfnamefont {X.}~\bibnamefont
  {Waintal}},\ }\bibfield  {title} {\enquote {\bibinfo {title} {{What Limits
  the Simulation of Quantum Computers?}}}\ }\href
  {https://link.aps.org/doi/10.1103/PhysRevX.10.041038} {\bibfield  {journal}
  {\bibinfo  {journal} {Phys. Rev. X}\ }\textbf {\bibinfo {volume} {10}},\
  \bibinfo {pages} {041038} (\bibinfo {year} {2020})}\BibitemShut {NoStop}%
\bibitem [{\citenamefont {{Stoudenmire}}\ and\ \citenamefont
  {{Waintal}}(2023)}]{Grover_no_quantum_advantage}%
  \BibitemOpen
  \bibfield  {author} {\bibinfo {author} {\bibfnamefont {E.~M.}\ \bibnamefont
  {{Stoudenmire}}}\ and\ \bibinfo {author} {\bibfnamefont {X.}~\bibnamefont
  {{Waintal}}},\ }\href@noop {} {\enquote {\bibinfo {title} {{Grover's
  Algorithm Offers No Quantum Advantage}},}\ } (\bibinfo {year} {2023}),\
  \Eprint {http://arxiv.org/abs/2303.11317} {arXiv:2303.11317} \BibitemShut
  {NoStop}%
\bibitem [{\citenamefont {Ayral}\ \emph {et~al.}(2023)\citenamefont {Ayral},
  \citenamefont {Louvet}, \citenamefont {Zhou}, \citenamefont {Lambert},
  \citenamefont {Stoudenmire},\ and\ \citenamefont {Waintal}}]{DMRG_QC}%
  \BibitemOpen
  \bibfield  {author} {\bibinfo {author} {\bibfnamefont {T.}~\bibnamefont
  {Ayral}}, \bibinfo {author} {\bibfnamefont {T.}~\bibnamefont {Louvet}},
  \bibinfo {author} {\bibfnamefont {Y.}~\bibnamefont {Zhou}}, \bibinfo {author}
  {\bibfnamefont {C.}~\bibnamefont {Lambert}}, \bibinfo {author} {\bibfnamefont
  {E.~M.}\ \bibnamefont {Stoudenmire}}, \ and\ \bibinfo {author} {\bibfnamefont
  {X.}~\bibnamefont {Waintal}},\ }\bibfield  {title} {\enquote {\bibinfo
  {title} {{Density-Matrix Renormalization Group Algorithm for Simulating
  Quantum Circuits with a Finite Fidelity}},}\ }\href
  {https://link.aps.org/doi/10.1103/PRXQuantum.4.020304} {\bibfield  {journal}
  {\bibinfo  {journal} {PRX Quantum}\ }\textbf {\bibinfo {volume} {4}},\
  \bibinfo {pages} {020304} (\bibinfo {year} {2023})}\BibitemShut {NoStop}%
\bibitem [{\citenamefont {Wang}\ \emph {et~al.}(2017)\citenamefont {Wang},
  \citenamefont {Hill},\ and\ \citenamefont {Hollenberg}}]{Shor_alg_MPS}%
  \BibitemOpen
  \bibfield  {author} {\bibinfo {author} {\bibfnamefont {D.~S.}\ \bibnamefont
  {Wang}}, \bibinfo {author} {\bibfnamefont {C.~D.}\ \bibnamefont {Hill}}, \
  and\ \bibinfo {author} {\bibfnamefont {L.~C.~L.}\ \bibnamefont
  {Hollenberg}},\ }\bibfield  {title} {\enquote {\bibinfo {title} {{Simulations
  of Shor's algorithm using matrix product states}},}\ }\href
  {https://doi.org/10.1007/s11128-017-1587-x} {\bibfield  {journal} {\bibinfo
  {journal} {Quant. Inf. Process.}\ }\textbf {\bibinfo {volume} {16}} (\bibinfo
  {year} {2017})}\BibitemShut {NoStop}%
\bibitem [{\citenamefont {Dang}\ \emph {et~al.}(2019)\citenamefont {Dang},
  \citenamefont {Hill},\ and\ \citenamefont {Hollenberg}}]{Shor_alg_MPS_optim}%
  \BibitemOpen
  \bibfield  {author} {\bibinfo {author} {\bibfnamefont {A.}~\bibnamefont
  {Dang}}, \bibinfo {author} {\bibfnamefont {C.~D.}\ \bibnamefont {Hill}}, \
  and\ \bibinfo {author} {\bibfnamefont {L.~C.~L.}\ \bibnamefont
  {Hollenberg}},\ }\bibfield  {title} {\enquote {\bibinfo {title} {Optimising
  {M}atrix {P}roduct {S}tate {S}imulations of {S}hor's {A}lgorithm},}\ }\href
  {\doibase 10.22331/q-2019-01-25-116} {\bibfield  {journal} {\bibinfo
  {journal} {{Quantum}}\ }\textbf {\bibinfo {volume} {3}},\ \bibinfo {pages}
  {116} (\bibinfo {year} {2019})}\BibitemShut {NoStop}%
\bibitem [{\citenamefont {Woolfe}\ \emph {et~al.}(2017)\citenamefont {Woolfe},
  \citenamefont {Hill},\ and\ \citenamefont {Hollenberg}}]{Woolfe2017}%
  \BibitemOpen
  \bibfield  {author} {\bibinfo {author} {\bibfnamefont {K.~J.}\ \bibnamefont
  {Woolfe}}, \bibinfo {author} {\bibfnamefont {C.~D.}\ \bibnamefont {Hill}}, \
  and\ \bibinfo {author} {\bibfnamefont {L.~C.~L.}\ \bibnamefont
  {Hollenberg}},\ }\bibfield  {title} {\enquote {\bibinfo {title} {{Scaling and
  efficient classical simulation of the quantum Fourier transform}},}\ }\href
  {\doibase 10.26421/qic17.1-2-1} {\bibfield  {journal} {\bibinfo  {journal}
  {Quant. Inf. Comput.}\ }\textbf {\bibinfo {volume} {17}},\ \bibinfo {pages}
  {1} (\bibinfo {year} {2017})}\BibitemShut {NoStop}%
\bibitem [{\citenamefont {Fishman}\ \emph {et~al.}(2022)\citenamefont
  {Fishman}, \citenamefont {White},\ and\ \citenamefont
  {Stoudenmire}}]{Itensors}%
  \BibitemOpen
  \bibfield  {author} {\bibinfo {author} {\bibfnamefont {M.}~\bibnamefont
  {Fishman}}, \bibinfo {author} {\bibfnamefont {S.~R.}\ \bibnamefont {White}},
  \ and\ \bibinfo {author} {\bibfnamefont {E.~M.}\ \bibnamefont
  {Stoudenmire}},\ }\bibfield  {title} {\enquote {\bibinfo {title} {{The
  ITensor Software Library for Tensor Network Calculations}},}\ }\href
  {https://scipost.org/10.21468/SciPostPhysCodeb.4} {\bibfield  {journal}
  {\bibinfo  {journal} {SciPost Phys. Codebases}\ ,\ \bibinfo {pages} {4}}
  (\bibinfo {year} {2022})}\BibitemShut {NoStop}%
\bibitem [{\citenamefont {{Niedermeier}}(2024)}]{QSim_github}%
  \BibitemOpen
  \bibfield  {author} {\bibinfo {author} {\bibfnamefont {M.}~\bibnamefont
  {{Niedermeier}}},\ }\href {https://github.com/MarcelNiedermeier/Quantunity}
  {\enquote {\bibinfo {title} {Quantunity},}\ } (\bibinfo {year} {2024}),\
  \bibinfo {note} {accessed: 2023-03-26}\BibitemShut {NoStop}%
\bibitem [{\citenamefont {Qi}\ \emph {et~al.}(2006)\citenamefont {Qi},
  \citenamefont {Wu},\ and\ \citenamefont {Zhang}}]{QWZ2006}%
  \BibitemOpen
  \bibfield  {author} {\bibinfo {author} {\bibfnamefont {X.-L.}\ \bibnamefont
  {Qi}}, \bibinfo {author} {\bibfnamefont {Y.-S.}\ \bibnamefont {Wu}}, \ and\
  \bibinfo {author} {\bibfnamefont {S.-C.}\ \bibnamefont {Zhang}},\ }\bibfield
  {title} {\enquote {\bibinfo {title} {{Topological quantization of the spin
  Hall effect in two-dimensional paramagnetic semiconductors}},}\ }\href
  {\doibase 10.1103/PhysRevB.74.085308} {\bibfield  {journal} {\bibinfo
  {journal} {Phys. Rev. B}\ }\textbf {\bibinfo {volume} {74}},\ \bibinfo
  {pages} {085308} (\bibinfo {year} {2006})}\BibitemShut {NoStop}%
\bibitem [{Note2()}]{Note2}%
  \BibitemOpen
  \bibinfo {note} {The probability density is computed as $ P(X_W) = \DOTSB
  \sum@ \slimits@ _j \protect \frac {\varepsilon }{\varepsilon ^2 + (X_W -
  X_W^j)^2}, $ where $X^j_W=\theta _B^j$ are the Berry phases obtained from the
  quantum phase estimation algorithm, and the small parameter $\varepsilon =
  0.1$ broadens the peaks.}\BibitemShut {Stop}%
\bibitem [{\citenamefont {Hatsugai}(2007)}]{Hatsugai2007}%
  \BibitemOpen
  \bibfield  {author} {\bibinfo {author} {\bibfnamefont {Y.}~\bibnamefont
  {Hatsugai}},\ }\bibfield  {title} {\enquote {\bibinfo {title} {{Quantized
  Berry phases for a local characterization of spin liquids in frustrated spin
  systems}},}\ }\href {http://dx.doi.org/10.1088/0953-8984/19/14/145209}
  {\bibfield  {journal} {\bibinfo  {journal} {J. Phys.: Condens. Matter}\
  }\textbf {\bibinfo {volume} {19}},\ \bibinfo {pages} {145209} (\bibinfo
  {year} {2007})}\BibitemShut {NoStop}%
\bibitem [{Hel(2024{\natexlab{b}})}]{Helmi_VTT}%
  \BibitemOpen
  \href {https://vttresearch.github.io/quantum-computer-documentation/helmi/}
  {\enquote {\bibinfo {title} {Helmi commercial documentation},}\ } (\bibinfo
  {year} {2024}{\natexlab{b}}),\ \bibinfo {note} {accessed:
  2024-03-19}\BibitemShut {NoStop}%
\end{thebibliography}%

%%%%%%%%%%%%%%
\end{document}